\documentclass[fleqn,usenatbib]{mnras}

\usepackage{newtxtext,newtxmath}
\usepackage[T1]{fontenc}

\DeclareRobustCommand{\VAN}[3]{#2}
\let\VANthebibliography\thebibliography
\def\thebibliography{\DeclareRobustCommand{\VAN}[3]{##3}\VANthebibliography}

\usepackage{graphicx}	% Including figure files
\usepackage{amsmath}	% Advanced maths commands
\usepackage{enumitem} 

\title[Massive Sgr and GD-1]{The impact of a massive Sagittarius dSph on GD-1-like streams}

\author[A. M. Dillamore et al.]{
Adam M. Dillamore,$^{1}$\thanks{E-mail: amd206@cam.ac.uk}
Vasily Belokurov,$^{1,2}$
N. Wyn Evans$^{1}$
and Adrian M. Price-Whelan$^{2}$
\\
$^{1}$Institute of Astronomy, University of Cambridge, Madingley Road, Cambridge CB3 0HA, UK\\
$^{2}$Center for Computational Astrophysics, Flatiron Institute, 162 5th Avenue, New York, NY 10010, USA
}

\date{Accepted XXX. Received YYY; in original form ZZZ}

\pubyear{2022}

\begin{document}
\label{firstpage}
\pagerange{\pageref{firstpage}--\pageref{lastpage}}
\maketitle

% Abstract of the paper
\begin{abstract}
We investigate the effects of a massive ($\gtrsim4\times10^{10}M_\odot$) Sagittarius dwarf spheroidal galaxy (Sgr) on stellar streams using test particle simulations in a realistic Milky Way potential. We find that Sgr can easily disrupt streams formed more than $\sim3$~Gyr ago, while stars stripped more recently are generally unaffected. In certain realizations, Sgr is able to produce asymmetry between the leading and trailing tails of Pal 5, qualitatively similar to observations. Using data from the \textit{Gaia} space telescope and elsewhere, we fit models to the GD-1 stream in the presence of a Sgr with various initial masses. While the best-fitting models do show perturbations resulting from interactions with Sgr, we find that the level of disruption is not significantly greater than in the observed stream. To investigate the general effects of Sgr on a population of streams, we generate 1000 mock streams on GD-1-like orbits with randomized orientations. Some streams show clear evidence of disruption, becoming folded on the sky or developing asymmetry betweeen their two tails. However, many survive unaffected and the peak surface brightness of stars is decreased by no more than $\sim0.3$~mag/arcsec$^2$ on average. We conclude that Sgr having an initial mass of $\gtrsim4\times10^{10}M_\odot$ is compatible with the survival and detection of streams formed more than 3~Gyr ago.
\end{abstract}

% Select between one and six entries from the list of approved keywords.
% Don't make up new ones.

%%%%%%%%%%%%%%%%%%%%%%%%%%%%%%%%%%%%%%%%%%%%%%%%%%

%%%%%%%%%%%%%%%%% BODY OF PAPER %%%%%%%%%%%%%%%%%%

\section{Introduction}

The Sagittarius dwarf spheroidal galaxy (Sgr), discovered by \citet{Ibata_95}, is one of the closest and brightest satellites of the Milky Way. Having just passed pericentre, it is currently undergoing strong tidal disruption and is expected to completely dissolve over the next billion years \citep{last_breath}. Stars have been stripped from Sgr to form the Sagittarius stream, a pair of long tails which lead and trail Sgr. These wrap around the Milky Way in a plane roughly perpendicular to its disc \citep[see e.g.][]{Ibata_01, Majewski, Belokurov_streams, Newberg}, and are useful probes of both the Milky Way's potential and the orbit of Sgr \citep[e.g.][]{Ibata_01, Fe06, Law, Gibbons, tango}. Dynamical modelling of the disruption of Sgr by \citet{Law} suggested that the present-day mass of the Sgr remnant is about $2.5\times10^8M_\odot$, and more recently \citet{last_breath} found a mass of $\sim4\times10^8M_\odot$. This is much less than the current mass of the Large Magellanic Cloud (LMC), about $1.4\times10^{11}M_\odot$ \citep{Erkal_Orphan}.

However, since Sgr has already been significantly disrupted, its initial mass must have been much higher. Using N-body models, \citet{Jiang} argued that the available data was compatible with a Sgr of initial mass $\sim10^{11}M_\odot$. \citet{Niederste-Ostholt} calculated the total luminosity of the Sgr debris and estimated the original mass of its dark matter (DM) halo to be $\sim10^{10}M_\odot$. Various lines of evidence have more recently emerged to suggest that this mass may have been $>10^{10}M_\odot$. \citet{Gibbons2017} used simulations of the Sgr stream in conjunction with chemistry and kinematics from Sloan Digital Sky Survey/SEGUE data \citep{SEGUE} to derive an initial mass estimate of $>6\times10^{10}M_\odot$. Using an abundance matching technique \citet{Read} found a similar value of $5\times10^{10}M_\odot$.

Further evidence may be provided from the discovery by \citet{Antoja} of a spiral pattern in the phase space of Milky Way disc stars using data from the \textit{Gaia} space telescope \citep{Gaia_DR2}. This is sometimes known as a `snail' or `phase space spiral'. Studies have shown that such a spiral can emerge from interaction with a disc-crossing dwarf galaxy \citep[e.g.][]{Antoja, Binney_spiral} such as Sgr, providing a possible indicator of its mass at earlier times. \citet{Laporte18} ran simulations of Sgr with initial mass $\sim10^{11}M_\odot$ and showed that many observed perturbations of the disc can be reproduced~\citep{Laporte19}. \citet{Bland-Hawthorn} have suggested that Sgr must have been losing mass at a rate of 0.5-1.0 dex per orbit for it to have excited the phase space spiral in the last 1-2 Gyr. However, \citet{Bennett} have argued that Sgr cannot be the sole creator of the perturbations, since their simulations were unable to match both the perturbations and the present mass of the Sgr remnant. The mass of Sgr at earlier times therefore remains an open question, one that is central to understanding the last few billion years of the Milky Way.

Stellar streams are produced by dissolving globular clusters or satellite galaxies. As the progenitor orbits the host galaxy, tidal forces strip stars from the Lagrange points to create one or two tails which lead or trail the progenitor on its orbit \citep[e.g.][]{Lynden-Bell_streams, Binney_Tremaine}. Cold stellar streams, formed from disrupted globular clusters, are narrow with low velocity dispersion. They provide a visible trace of an approximate orbit through the galaxy, allowing the potential of the Milky Way to be constrained \citep{Johnston, Bonaca_information}. Perturbations induced by encounters with DM subhaloes are expected to heat and create gaps in cold stellar streams \citep[see e.g.][]{Ibata_02, Erkal_gaps15, Erkal_gaps16} which can be compared with predictions from $\Lambda$CDM cosmology. Encounters with known satellite galaxies can also perturb streams. For example, the Orphan stream \citep{Grillmair_Orphan,Belokurov_Orphan} was found by \citet{Koposov_Orphan} to contain a misalignment between its on-sky track and the proper motions of its stars. \citet{Erkal_Orphan} argued that this arose from an interaction with the LMC, and by modelling this interaction were able to constrain both the Milky Way potential and the mass of the LMC. Streams may also be perturbed by interactions with the galactic bar \citep{Pearson}, and resonances arising from the Milky Way's potential \citep{Yavetz_2020, Yavetz_2022}.

The GD-1 stream was discovered by \citet{Grillmair} using Sloan Digital Sky Survey (SDSS) data. GD-1 is cold and narrow, and spans over $100^\circ$ on the sky \citep{Webb}. Excellent 6D data is available for the stream \citep[e.g.][]{Koposov_GD1, Price-Whelan_GD1, deBoer18, deBoer20}, allowing relatively tight constraints on its orbit. Various studies \citep[e.g.][]{Bowden, Malhan_potential} have utilised this to constrain the potential of the Milky Way.

The pericentric and apocentric distances of GD-1 are around 14~kpc \citep{Bowden} and 26~kpc \citep{Koposov_GD1} respectively. The radial range explored by GD-1 therefore overlaps with that of Sgr, which reached a radius of $r\approx16$~kpc at its last pericentre \citep{last_breath}. \citet{deBoer20} showed that an interaction between GD-1 and Sgr $\approx 3$~Gyr ago could reproduce some of the off-track features of GD-1. These include the `spur' and the `blob' \citep{Price-Whelan_GD1}, which lie $\sim1^\circ$ away from the main track of the stream on the sky. The stream also has wiggles and density variations \citep[e.g.][]{deBoer18, deBoer20}. All these features are thought to arise from encounters with DM subhaloes \citep{Bonaca_GD1}, and may provide information about the distributions and nature of these subhaloes in the Milky Way. It has also been proposed by \citet{Malhan_cocoon} and \citet{Qian} that the off-track features could arise if GD-1 was formed from a cluster accreted from a satellite galaxy. In these scenarios, the off-track features are comprised of stars stripped from the cluster before it was accreted into the Milky Way.

Like many other Galactic stellar streams \citep[e.g. the Orphan stream,][]{Grillmair_Orphan, Belokurov_streams,Koposov_ATLAS,Shipp_DES}, no progenitor cluster has been discovered for GD-1, although \citet{Webb} have suggested possible scenarios for the dissolution and location of the progenitor. One possibility is that the cluster dissolved more than 2.5~Gyr ago, in which case the stream would have had ample time to interact with subhaloes including Sgr.

In this study, we investigate the effects of a Sgr of mass $>10^{10}M_\odot$ on stellar streams in the Milky Way. We focus on GD-1 due to the possibility of encounters with Sgr, and the high-quality data available over a large expanse of sky. While uncertainties in the Milky Way's potential and the position of Sgr are far too great to predict or model the nature of such encounters, the masses of Sgr being considered mean that even distant encounters may cause significant disruption to the stream. It is therefore conceivable that such disruption is unavoidable for streams orbiting at similar radii to the pericentre of Sgr. We seek to answer whether this is the case, and whether the observed GD-1 stream is compatible with it having survived encounters with a massive Sgr. If not, this would suggest that observed streams like GD-1 must have formed after Sgr had lost much of its initial mass. This would have consequences for future studies of GD-1-like streams.

The paper is arranged as follows. In Section~\ref{section:Setup} we describe the simulation setup, including the models for Sgr and the method for generating streams using the Lagrange Cloud Stripping technique. We then test these models on mock streams generated from known globular clusters in Section~\ref{section:qualitative}, before fitting models to data from the GD-1 stream in Section~\ref{section:fitting}. In Section~\ref{section:survival}, we generate 1000 streams on GD-1-like orbits to investigate whether a population of streams can survive the passage of a massive Sgr. Finally we summarise our findings in Section~\ref{section:conclusions}.

\section{Setup}\label{section:Setup}
We use the package \texttt{AGAMA} \citep{agama} to conduct test particle simulations of stellar streams in the combined potential of the Milky Way (MW), Large Magellanic Cloud (LMC) and Sagittarius Dwarf Galaxy (Sgr). Throughout this paper we use a right-handed Galactocentric coordinate system $(x, y, z)$, where the Sun is situated at $(-8.1, 0, 0.02)$~kpc with a velocity of $(12.9, 245.6, 7.8)$~km/s, consistent with \texttt{ASTROPY} \citep{astropy:2018}.

\subsection{Potential}

In setting up the potential of the Milky Way and the LMC, we closely follow the methods described in \citet{tango}, where a detailed description can be found.

\subsubsection{Milky Way}

For the MW's potential, we choose the triaxial fiducial model used by \citet{tango} to model the Sgr stream. Its halo has a radius-dependent shape: the inner halo is axisymmetric and flattened perpendicular to the disc, while the outer halo ($\gtrsim50$~kpc) is triaxial, with its major axis perpendicular to the disc. This potential was shown to produce reasonable matches to the Sgr stream, and is therefore an appropriate choice for studying dynamics in its vicinity. The potential is described in more detail in Appendix~\ref{section:appendix_MW}, and also Section 3.4 and Table 1 in \citet{tango}.

\subsubsection{LMC}

The contribution from the LMC to our potential consists of two components. In addition to the direct gravitational potential, there is a uniform (but time-varying) acceleration due to the reflex motion of the Milky Way's centre of mass towards the LMC. This correction was found by \citet{Gomez} to have a significant effect on the phase space structure of the Sgr debris, and \citet{tango} concluded that the recoil was necessary to find a satisfactory fit to the stream.

The LMC is modelled with a spherically symmetric density profile
\begin{equation}
    \rho_\mathrm{LMC}\propto r^{-1}(1+r/r_\mathrm{scale})^{-2}\mathrm{exp}[-(r/r_\mathrm{trunc})^2],
\end{equation}
which is a truncated Navarro-Frenk-White (NFW) model. We set the total mass to be $M_\mathrm{LMC}=1.5\times10^{11}M_\odot$, with a scale radius $r_\mathrm{scale}\approx15.6$~kpc and truncation radius $r_\mathrm{trunc}=10r_\mathrm{scale}\approx156$~kpc. The relation between $M_\mathrm{LMC}$ and $r_\mathrm{scale}$ was chosen by \citet{tango} to satisfy observational constraints on the LMC's circular velocity.

The motions of the MW and LMC under their mutual gravitational attraction are found by numerically integrating the equations
\begin{equation}\label{eq:LMC_DEs}
\begin{split}
    \dot{\mathbf{x}}_\mathrm{MW}&=\mathbf{v}_\mathrm{MW}\\
    \dot{\mathbf{v}}_\mathrm{MW}&=-\nabla\Phi_\mathrm{LMC}(\mathbf{x}_\mathrm{MW}-\mathbf{x}_\mathrm{LMC})+\tilde{\mathbf{a}}_\mathrm{MW}\\
    \dot{\mathbf{x}}_\mathrm{LMC}&=\mathbf{v}_\mathrm{LMC}\\
    \dot{\mathbf{v}}_\mathrm{LMC}&=-\nabla\Phi_\mathrm{MW}(\mathbf{x}_\mathrm{LMC}-\mathbf{x}_\mathrm{MW})+\tilde{\mathbf{a}}_\mathrm{LMC},
\end{split}
\end{equation}
where $\mathbf{x}_\mathrm{MW}$ and $\mathbf{x}_\mathrm{LMC}$ ($\mathbf{v}_\mathrm{MW}$ and $\mathbf{v}_\mathrm{LMC}$) are the position vectors (velocities) of the MW and LMC in an inertial frame, $\Phi_\mathrm{MW}$ and $\Phi_\mathrm{LMC}$ are the gravitational potentials, and $\tilde{\mathbf{a}}_\mathrm{MW}$ and $\tilde{\mathbf{a}}_\mathrm{LMC}$ are correction terms to account for the effects of dynamical friction and deformation. These are calculated by calibration with the N-body simulations of \citet{tango}.

We use the same present-day phase space position for the centre of the LMC as \citet{tango}. The R.A. and declination are $\alpha_\mathrm{LMC}=81^\circ$ and $\delta_\mathrm{LMC}=-69.75^\circ$ respectively, taken from \citet{vasiliev_LMC}, with corresponding proper motions $\mu_\mathrm{\alpha^*,LMC}=1.8$~mas/yr and $\mu_\mathrm{\delta,LMC}=0.35$~mas/yr. The distance and radial velocity are 50~kpc \citep{Freedman} and 260~km/s \citet{vdMarel}.

Eqns~(\ref{eq:LMC_DEs}) were integrated backwards from the present-day ($t=0$) to an initial time $t_\mathrm{start}=-6$~Gyr. The trajectory of the MW was used to compute the resulting uniform acceleration potential, as experienced by a particle at rest relative to the MW. This was added to the MW and LMC gravitational potentials in the non-inertial galactocentric coordinate system to obtain a combined MW-LMC potential consistent with the motion of the MW.
\begin{figure}
    \centering
    \includegraphics[width=\columnwidth]{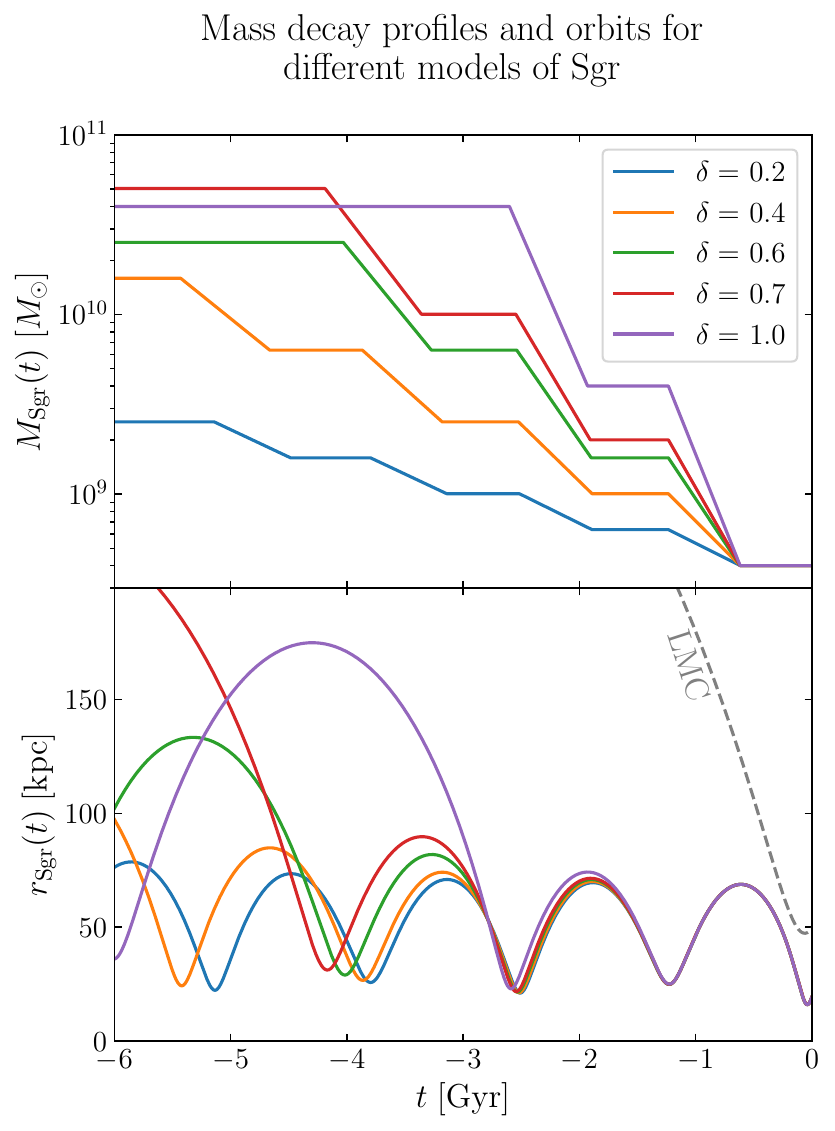}
    \caption{\textbf{Top panel}: masses of the different Sgr models as a function of time. The periods of exponentially decreasing mass (linear on the logarithmic plot) occur between pericentres and subsequent apocentres. All models have $M_\mathrm{Sgr}=4\times10^8M_\odot$ at the present-day. \textbf{Bottom panel}: Galactocentric distances of Sgr models as a function of time, with the LMC included for comparison. The models all follow similar orbits over their last two periods, and have a pericentre of $r_\mathrm{Sgr}\approx20$~kpc at $t\approx-2.7$~Gyr. Before this time, the more massive models ($\delta\gtrsim0.5$) are on longer-period orbits with larger apocentres, before dynamical friction causes their orbits to decay.}
    \label{fig:Sgr_mass_orbits}
\end{figure}
\begin{figure*}
    \centering
    \includegraphics[width=\textwidth]{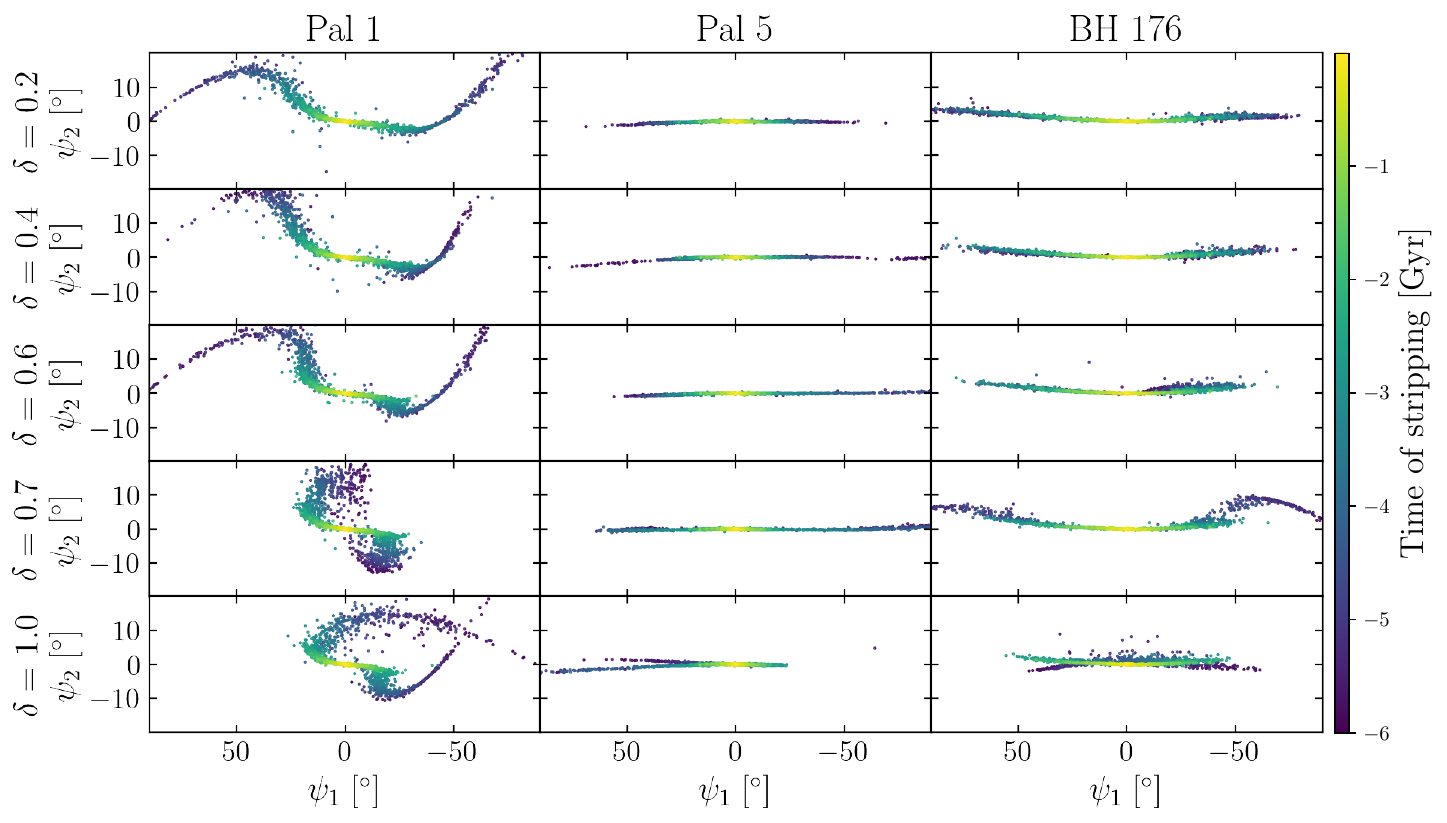}
    \caption{Mock stellar streams generated from three globular clusters, with five different values of the Sgr mass decay rate parameter $\delta$. All panels show the streams as viewed on the sky from the galactic centre, in coordinate systems where the $\psi_1$ axis is aligned with the instantaneous orbital plane of the progenitor cluster (situated at $\psi_1=\psi_2=0$). The progenitor's motion is towards the left (positive $\psi_1$). Both leading and trailing streams are shown, and the stars are colour-coded by the time at which they were stripped from the progenitor; purple is earliest and yellow is most recent. In all three cases increasing the mass of Sgr increases the disruption of streams. The stars stripped earliest (at $t\lesssim-3$~Gyr) are most affected, while the more recently stripped stars are able to form narrow tracks with even the highest masses of Sgr. An animated version of this figure can be viewed at \url{https://www.youtube.com/playlist?list=PLEleLLhXAwEMx6GcSror-iF-QsskHrXYM}.}
    \label{fig:GC_streams}
\end{figure*}

\subsubsection{Sgr}\label{section:setup_Sgr}

We model the Sgr dSph as a Hernquist sphere with time-varying mass and scale radius. Again following \citet{tango}, we place the present-day position of the Sgr remnant at $\alpha_\mathrm{Sgr}=283.76^\circ$, $\delta_\mathrm{Sgr}=-30.48^\circ$, with proper motions $\mu_{\alpha^*,Sgr}=-2.7$~mas/yr and $\mu_{\delta,Sgr}=-1.35$~mas/yr. The distance and radial velocity are 27~kpc and 142~km/s. This choice of distance was found by \citet{tango} to provide better fits to the Sgr stream than other values \citep[e.g. 26.5~kpc,][]{last_breath}.

We integrate the orbit of Sgr in the combined MW-LMC-reflex potential, including an acceleration due to dynamical friction 
\citep{Chandrasekhar, tango}:
\begin{equation}\label{eq:DF}
\begin{split}
    \mathbf{a}_\mathrm{DF}&=-4\pi\rho\,\mathrm{ln} \Lambda \, G M_\mathrm{Sgr}\,\frac{ \mathbf{v}_\mathrm{Sgr}}{|v_\mathrm{Sgr}|^3}[\mathrm{erf}(X)-2\pi^{-1/2}X\mathrm{exp}(-X^2)],\\
    X&\equiv\frac{|v_\mathrm{Sgr}|}{\sqrt{2}\sigma}.
\end{split}
\end{equation}
Here $\mathbf{v}_\mathrm{Sgr}$ is the velocity of Sgr in the Galactocentric frame, $\rho$ is the mass density of the MW (and LMC) at the location of Sgr, $\sigma$ is the velocity dispersion of particles at that location, and $\mathrm{ln}\Lambda$ is the Coulomb logarithm. While $\rho$ is calculated from the MW+LMC potential, $\sigma$ and $\mathrm{ln}\Lambda$ must be prescribed. Following \citet{tango}, we set $\sigma=120$~km/s and $\mathrm{ln}\Lambda=3$.

A crucial part of this work is the choice of mass decay profile $M_\mathrm{Sgr}(t)$ of Sgr as it orbits the Milky Way. We wish for $M_\mathrm{Sgr}(t)$ to be continuous, realistic and consistent with eqn~(\ref{eq:DF}), while allowing us to consider a high initial mass ($>10^{10}M_\odot$). We make the assumption that the total mass decreases by a factor of $10^\delta$ per pericentre passage. The `decay rate parameter' $\delta>0$ is the logarithmic mass lost per orbit in dex; i.e. $\mathrm{log}_{10}M_\mathrm{Sgr}$ decreases by $\delta$ every orbit. For continuity of $M_\mathrm{Sgr}(t)$, the mass is decreased exponentially between each pericentre and the subsequent apocentre (such that $\mathrm{log}\,M_\mathrm{Sgr}(t)$ decreases uniformly). It is then held constant until the next pericentre. Since only a small fraction of  an orbital period has elapsed since the most recent pericentre passage, we do not remove any mass between this time and the present-day. The present-day mass is set to $M_\mathrm{Sgr}(0)=4\times10^8M_\odot$, the total mass of the Sgr remnant found by \citet{last_breath}. We assume that the Hernquist radius scales as $M_\mathrm{Sgr}^{1/3}(t)$ and set the present-day value to 1.4~kpc. This gives a peak circular velocity of 17.5~km/s, similar to the best-fitting models of Sgr in \citet{last_breath}. This prescription would also give a scale radius of 8.2~kpc for a model of total mass $8\times10^{10}M_\odot$, similar to the L2 model of \citet{Laporte18}.

We approximate Sgr as a test particle moving in the MW potential, since simulating the mutual interaction of Sgr and the MW would require additional assumptions about how the mass stripped from Sgr is distributed in the MW (see Appendix~\ref{section:Sgr_orbit} for a justification of this approximation). However, we do include the uniform acceleration potential resulting from the MW's acceleration towards Sgr. Since we are considering masses of Sgr up to $\sim1/3$ that of the LMC, this effect is non-negligible. This is especially important when considering perturbations on streams; the MW reflex acceleration causes partial cancellation of Sgr-induced perturbations, so neglecting this effect would lead to overestimation of these perturbations. We include this reflex in the same manner as that due to the LMC; we calculate the Sgr-centric acceleration evaluated at the centre of the MW, and subtract this acceleration from all equations of motion in the galactocentric frame. The LMC is ignored in this calculation because it is far from the MW at early times, when Sgr has non-negligible mass.

For the analysis we consider up to five models with different decay rates, with \hbox{$\delta\in\{0.2, 0.4, 0.6, 0.7, 1.0\}$.} $M_\mathrm{Sgr}(t)$ and the galactocentric distance $r_\mathrm{Sgr}(t)$ are plotted in Fig.~\ref{fig:Sgr_mass_orbits} for these five models. The two models with lowest $\delta$ (0.2 and 0.4) do not experience significant orbital decay due to dynamical friction, and each have 5 pericentre passages during the $6$~Gyr. The $\delta=0.6$ and 0.7 models are similar to each other, both starting at large radii and experiencing 4 pericentre passages. The fastest-decaying model ($\delta=1.0$) only passes pericentre 3 times, the earliest being at $t\approx-2.8$~Gyr. As a result, the initial mass ($4\times10^{10}M_\odot$) is actually lower than that of the $\delta=0.7$ model ($5\times10^{10}M_\odot$). Care is therefore needed when interpreting our results in relation to the mass of Sgr at earlier times. The $\delta=0.7$ and $\delta=1.0$ models have initial masses somewhat less than those of \citet{Laporte18}. However, those authors' values of $8-14\times10^{10}M_\odot$ were the masses at infall, so their models would be more comparable to ours by the time of the first pericentric passage due to tidal stripping.

\subsection{Stream generation}

To rapidly generate realistic models of stellar streams we use the `Lagrange point stripping' technique outlined by \citet{Bowden}, also known as `modified Lagrange Cloud Stripping' \citep{Gibbons}. This is similar (though not identical) to the methods of \citet{Varghese} and \citet{Kupper}. We treat the stars in the stream as test particles moving in the combined potential of the MW, LMC, Sgr and a stream progenitor. This technique has been shown to produce streams which are indistinguishable from those in N-body simulations, while being much cheaper computationally to run \citep{Gibbons}.

The stream progenitor is modelled as a Plummer sphere of mass $M_\mathrm{prog}(t)$ and constant scale radius $a_\mathrm{prog}$. Its orbit is found by integrating back in time from some present-day phase space coordinates, through the combined MW+LMC+Sgr potential $\Phi$. At each time $t$ the locations of the Lagrange points are estimated as being along the line from the centre of the MW to the progenitor, at a displacement from the progenitor $\Delta r=\pm r_\mathrm{t}$. The tidal radius $r_\mathrm{t}$ is given by \citep{Gibbons, Bowden}
\begin{equation}
    r_\mathrm{t} = \left(\frac{GM_\mathrm{prog}}{\Omega^2-\frac{\mathrm{d}^2\Phi}{\mathrm{d}r^2}}\right)^{1/3},
\end{equation}
where $\Omega$ is the instantaneous angular speed of the progenitor about the MW's centre. Following \citet{Bowden}, we release particles from Galactocentric radii $r_\mathrm{strip}=r_\mathrm{prog}\pm\lambda r_\mathrm{t}$ where $\lambda=1.2$, rather than from the Lagrange points themselves. This ensures that a large majority of particles escape the progenitor and enter the streams.

The velocities of the stream particles are randomly drawn from a 3D isotropic Gaussian centred on $\mathbf{v}_\mathrm{strip}$ with standard deviation $\sigma_\mathrm{strip}$. The radial component of $\mathbf{v}_\mathrm{strip}$ is equal to that of the progenitor, while the tangential components are set equal to those of the point halfway between the centre of the progenitor and the Lagrange point. This choice has been used successfully by \citet{Kupper} and \citet{Bowden}.

We set the initial progenitor mass (at $t=-6$~Gyr) to $M_\mathrm{prog}=2\times10^4M_\odot$, and either keep $M_\mathrm{prog}$ constant or decrease it uniformly to zero at some time of dispersal $t_\mathrm{disp}$. The Plummer scale radius is set to $a_\mathrm{prog}=2$~pc, and the stripping velocity dispersion is $\sigma_\mathrm{strip}=0.5$~km/s. The total stellar mass of the visible GD-1 stream was estimated by \citet{deBoer20} to be $1.8\times10^4M_\odot$, so our choice of initial $M_\mathrm{prog}$ is above the absolute lower bound for the initial mass of the GD-1 progenitor. We later check whether using a lower mass affects our results (see Section~\ref{section:GD1_results}). We choose the velocity dispersion to obtain a model stream with a similar on-sky width to GD-1 (in the absence of strong perturbations).

\section{Qualitative results from mock streams}\label{section:qualitative}

\subsection{Generation of mock streams from known globular clusters}

We first test our models of a massive Sgr with mock streams orbiting at similar radii to GD-1, which in our potential is in the range $12\lesssim r/\mathrm{kpc}\lesssim 24$ (with no Sgr). Since we wish the progenitors of these streams to be on realistic globular cluster (GC) orbits, we search for suitable GCs in the catalogue by \citet{Vasiliev_GCs} based on \textit{Gaia} DR2 data.

We load the present-day 6D phase space positions of all the GCs in \citet{Vasiliev_GCs} via the \texttt{galpy} module \citep{galpy}. We integrate the orbits backwards to $t=-6$~Gyr in the MW+LMC potential, but with no Sgr. We find the maximum and minimum radii $r_\mathrm{max}$ and $r_\mathrm{min}$ of all orbits, and select those which satisfy $r_\mathrm{max}<25$~kpc and $r_\mathrm{min}>11$~kpc. This limits the GC orbits to the spherical shell explored by GD-1.

Of the 150 GCs in the catalogue, only 3 have orbits that satisfy these constraints. These are Pal 1, Pal 5 and BH 176. The latter's identity as a globular cluster is not agreed upon, and it has also been identified as an open cluster \citep{vdBergh, BH176}. Interestingly, Pal 5 has its own tidal tails \citep{Odenkirchen} which are among the best-studied of all cold stellar streams in the halo \citep[see e.g.][]{Odenkirchen,Rockosi2002,Dehnen2004,Grillmair2006,Kupper2015,Ibata2016,Erkal2017,Bonaca_Pal5}. We therefore take this opportunity to investigate models of the Pal 5 tails under the influence of a massive Sgr.

We use our Lagrange point stripping method to generate stellar streams from the Lagrange points of these GCs, keeping $M_\mathrm{prog}$ constant (i.e. the clusters do not disperse). We strip stars from both the inner and outer Lagrange points, generating leading and trailing tails respectively. We repeat for different choices of the Sgr mass decay rate parameter $\delta$. To display the streams, we calculate the instantaneous orbital plane of the progenitor as a function of time, and use this as the equator of a Galactocentric polar coordinate system with longitude $\psi_1$ and latitude $\psi_2$. We place the progenitor at $\psi_1=\psi_2=0$. The present-day appearances of the 3 streams are shown in Fig~\ref{fig:GC_streams} for various values of $\delta$. Along with Figs.~\ref{fig:gal_stills} and \ref{fig:stills}, an animated version can be viewed at \url{https://www.youtube.com/playlist?list=PLEleLLhXAwEMx6GcSror-iF-QsskHrXYM}.

Fig.~\ref{fig:GC_streams} demonstrates that a massive Sgr can have a highly disruptive effect on streams. While these effects vary between different cases, the structures of the streams formed at early times are all heavily affected by increasing the value of $\delta$. For $\delta=1.0$, the stars from Pal 1 stripped before $t\approx-3$~Gyr no longer form narrow tracks on the sky, and the two arms are twisted and significantly widened. The early-stripped stars from BH 176 are in a narrow track, but this is offset from the younger stream attached to the cluster. At lower values of $\delta$ the damage is a little less dramatic, but is still clearly visible at $\delta=0.6$, corresponding to a maximum Sgr mass of $2.5\times10^{10}M_\odot$. However, the disruption does not extend to stars stripped later (at $t>-3$~Gyr). All the panels in Fig.~\ref{fig:GC_streams} do show narrow streams comprised of recently stripped stars along the $\psi_1$ axes, albeit sometimes shorter than the full streams with smaller $\delta$. This is unsurprising, since encounters with a massive Sgr ($M_\mathrm{Sgr}\gtrsim10^{10}M_\odot$) can occur no later than $t\approx-2.7$~Gyr, around its third-from-last pericentre. This allows any material stripped after this time to form narrow streams, unaffected by Sgr. The Pal 1 and BH 176 stream models are comparable to those of \citet{Woudenberg}, who simulated the Jhelum stream in the presence of Sgr. They similarly showed that interactions with Sgr can result in a thin stream running parallel to older, more diffuse ones. This recreated some of the features of the observed Jhelum stream.

Pal 5 does not show extreme disruption to the same extent as the other two streams. However, for some values of $\delta$ there is strong asymmetry between the leading and trailing tails. For $\delta=0.6$ and $\delta=1.0$ in particular, one of the tails is significantly shorter than the other. This is interesting because the real Pal 5 stream does show such asymmetry; the leading tail is shorter than the trailing tail \citep{Dehnen2004,Erkal2017,Starkman, Bonaca_Pal5}. We discuss the asymmetry of our Pal 5 models further in Section~\ref{section:Pal5}.
\begin{figure*}
    \centering
    \includegraphics[width=\textwidth]{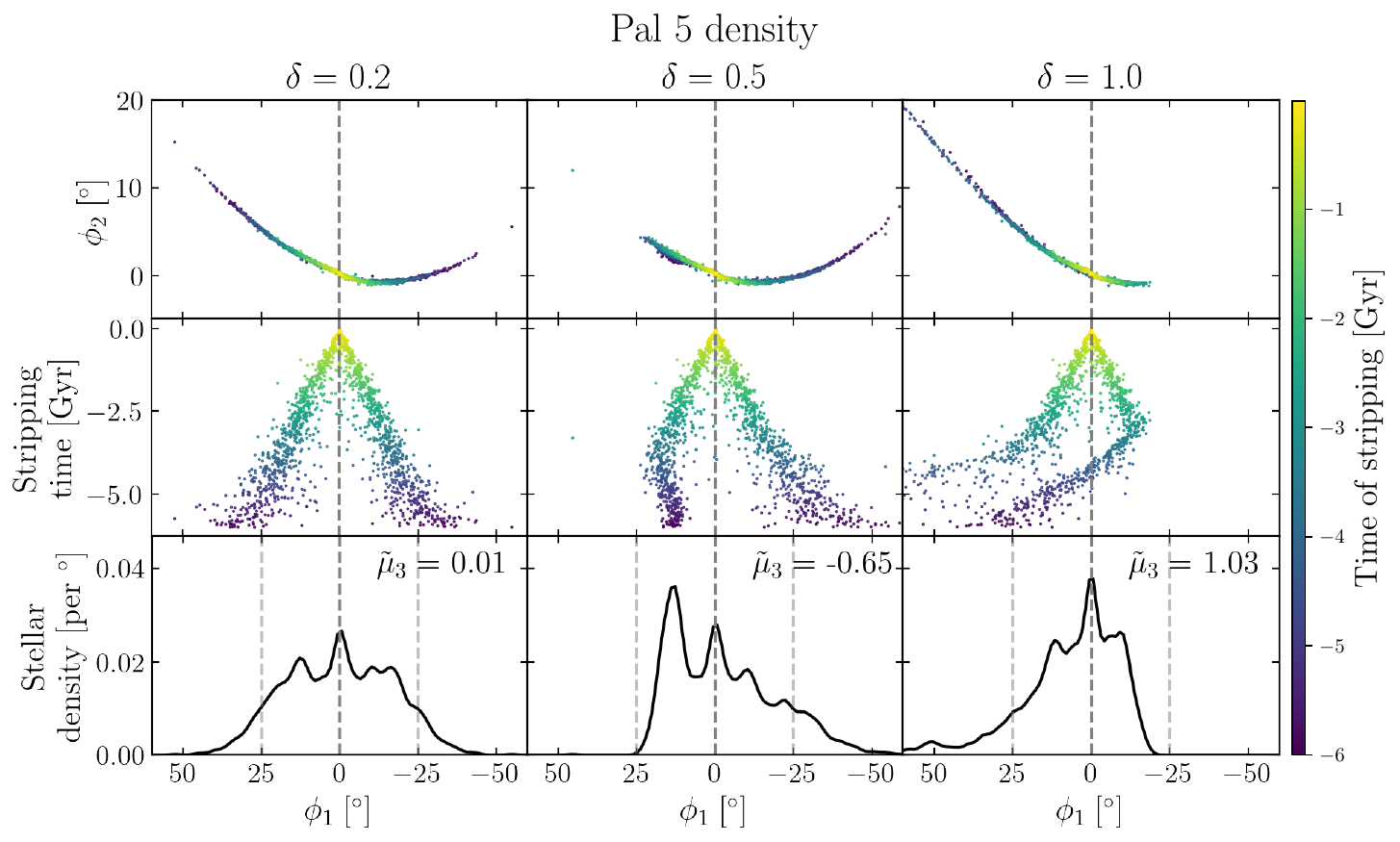}
    \caption{\textbf{Top row}: On-sky appearance of the mock Pal 5 streams in heliocentric stream coordinates ($\phi_1$, $\phi_2$), for three different choices of the Sgr mass decay rate parameter $\delta$. Stripping times of the stars are colour-coded. \textbf{Middle row}: The same stripping times plotted against $\phi_1$. \textbf{Bottom row}: kernel density estimates (KDEs) of the $\phi_1$ distributions of stars, with the skewness parameter $\tilde{\mu}_3$ shown. The progenitor cluster is at $\phi_1=0$ (marked by grey dashed lines) and its motion is in the direction of positive $\phi_1$. While the $\delta=0.2$ stream is close to symmetrical, in the other two cases there is asymmetry in the lengths and densities of two tails.}
    \label{fig:Pal5}
\end{figure*}
\subsection{The asymmetry of Pal 5}
\label{section:Pal5}

To compare our Pal 5 models to observations of the stream, we transform to a heliocentric coordinate system ($\phi_1$, $\phi_2$) approximately aligned with the stream. This is the \texttt{Pal5PriceWhelan18} coordinate system in the \texttt{gala} package \citep{gala} devised to study the Pal 5 stream \citep[e.g][]{Bonaca_Pal5, Price-Whelan_Pal5}. We repeat this for $\delta\in\{0.2, 0.5, 1.0\}$. The value $\delta=0.5$ corresponds to an initial Sgr mass of $4\times10^{10}M_\odot$, and was selected because it produces a good example of an asymmetric stream. We emphasise that these models each result from a single realisation of the progenitor positions, so represent only one set of possible outcomes. Fig.~\ref{fig:Pal5} shows the stream's on-sky appearance for each value of $\delta$, again with stripping times colour-coded. These times are also plotted against $\phi_1$ in the second row. To demonstrate the asymmetry about $\phi_1=0$, we calculate kernel density estimates (KDEs) of the $\phi_1$ distributions, first excluding stars outside the on-sky region $-60^\circ<\phi_1<60^\circ$, $-5^\circ<\phi_2<20^\circ$. As a measure of the asymmetry, we also compute skewness values $\tilde{\mu}_3$ of the same $\phi_1$ distributions. The KDEs and skewness values are shown in the bottom row of Fig.~\ref{fig:Pal5}.

With the low-mass Sgr ($\delta=0.2$), the stream is close to symmetric about the progenitor ($\phi_1=0$); the leading and trailing tails have a very similar length and density distribution. In the middle row, we see that the distance of stars from the progenitor in $\phi_1$ roughly correlates with the time since they were stripped (although with considerable scatter).

The symmetry is broken in the other two cases. With $\delta=0.5$, the trailing tail (at $\phi_1<0^\circ$, on the right-hand side) extends further than $50^\circ$ from the progenitor, while the leading tail (at $\phi_1>0^\circ$) is cut off sharply at $\phi_1\approx25^\circ$. The middle row reveals the structural reason for this: the leading tail is compressed so that the early-stripped stars lie closer to the progenitor. The KDE confirms that there is a significant enhancement in density at $\phi_1\approx15^\circ$. Since this compressed part is not perfectly aligned with the rest of the tail, it also results in an increased width along this part of the stream by almost a factor of 2.

The $\delta=1.0$ stream is superficially similar, though reversed; in this case, the leading stream extends beyond $50^\circ$ while the trailing stream is sharply cut off at $\phi_1\approx-25^\circ$. However, the stream is structurally different: the shortened tail does not show the same buildup of early-stripped stars or the resulting density peak. A dog-leg in the stripping time plot at $t\approx-2.7$~Gyr hints that the trailing stream has been `folded' and extends towards positive $\phi_1$, lying on top of the leading stream. This is confirmed by animations. After the Sgr pericentre at $t\approx-2.7$~Gyr, the trailing stream is twisted and its stars begin to overtake the progenitor, sometimes lying along the leading stream.

The observed tails of Pal 5 have complex structure which has been studied in detail by \citet{Erkal2017} and \citet{Bonaca_Pal5}. The leading tail is significantly shorter than the trailing one (although discoveries by \citet{Starkman} of possible extensions to the stream may dispute this). The leading tail also fans out before it is curtailed, and both tails contain wiggles and gaps (i.e. underdensities). The mock Pal 5 streams shown in Figure~\ref{fig:Pal5} exhibit some qualitative similarities to these features, most clearly the asymmetry in lengths. The stellar density plots on the bottom row also show underdensities a few degrees away from the cluster, even in the $\delta=0.2$ case. These are comparable to the gaps of Pal 5 shown in Fig. 5 of \citet{Erkal2017} and in Fig. 3 of \citet{Bonaca_Pal5}, though much less prominent. The greater width of the leading tail in the $\delta=0.5$ model is also similar to the `fan' of the real stream. Slightly different initial conditions can result in this `folded' tail becoming more misaligned with itself, resulting in an even larger apparent width on the sky. This type of perturbation is therefore able to reproduce both shortening and widening of the tail, at least qualitatively.

However, the scale of the asymmetry is considerably larger in our mock streams. While the shorter tails are both sharply cut off $25^\circ$ from the cluster, the observed leading tail of Pal 5 extends to only $\phi_1\approx7^\circ$. The second row of Fig.~\ref{fig:Pal5} shows that the stars stripped since $t\approx-2.7$~Gyr form a tail which is unaffected by Sgr, and extends to its full unperturbed length. Only the earlier-stripped stars are shifted in $\phi_1$. The cut-off angle of $25^\circ$ is therefore set by how far the more recently stripped stars drift from the cluster in the time since the interaction with Sgr. If the asymmetry of Pal 5 was caused solely by Sgr, this rate of drift must have been considerably slower than in our models. We checked whether changing the model parameters could achieve this, and found that a progenitor mass of $1\times10^4M_\odot$ reduced the cut-off angle to less than $15^\circ$ if we set $\lambda=0.5$ (i.e. particles are released halfway between the cluster centre and the Lagrange point). This mass is consistent with the value of $(1.39\pm0.65)\times10^4M_\odot$ reported by \citet{Baumgardt}. It may therefore be possible for some realistic model to match the observations even closer. Hence, we cannot rule out Sgr as the cause of the asymmetry in Pal 5, and this provides an alternative to other possible causes of the asymmetry, such as the rotation of the galactic bar \citep{Pearson}.

\section{Fitting the GD-1 stream}\label{section:fitting}

In this section, we investigate the effect of a massive Sgr on the GD-1 stream, which may have been able to interact with Sgr $\sim3$~Gyr ago. Such an interaction may have created observed off-track features such as the `spur' \citep{deBoer20}. We now take this idea further by questioning whether the present-day appearance of GD-1 is consistent with it having survived a pericentric passage of a more massive Sgr. To achieve this, we find maximum likelihood models of GD-1 constructed using the Lagrange cloud stripping technique.

\subsection{Data}
\begin{figure}
    \centering
    \includegraphics[width=0.9\columnwidth]{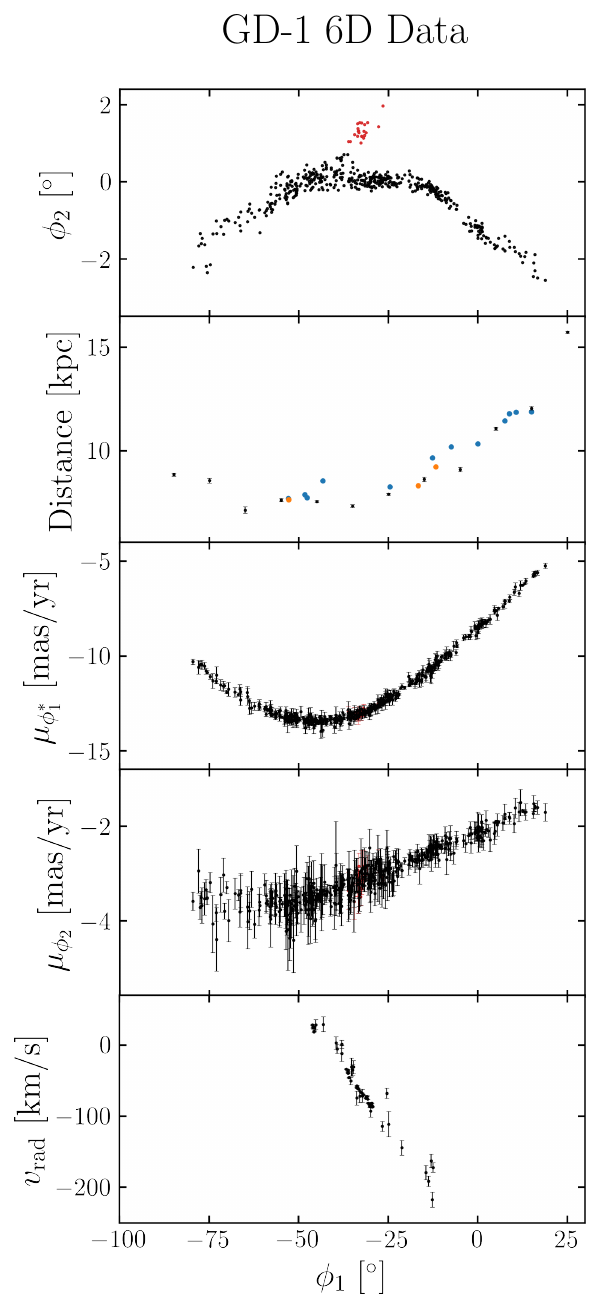}
    \caption{6D phase space data used for fitting the GD-1 stream (in black), with the excluded spur in red. From top to bottom, the coordinates are $\phi_2$, heliocentric distance, proper motions in $\phi_1$ and $\phi_2$, and heliocentric radial velocity. All are plotted against the stream coordinate $\phi_1$. The $\phi_1$ proper motion $\mu_{\phi_1^*}$ includes the $\mathrm{cos}\phi_2$ correction, and the velocities are not corrected for solar motion. The orange (blue) points in the distance panel represent the RRL (BHB) measurements, which are used to check the distance scale but not included in the likelihood calculations.}
    \label{fig:data}
\end{figure}

\subsubsection{4D \textit{Gaia} data}

We use astrometric data from the \textit{Gaia} mission's Early Data Release 3 (EDR3; \citealt{Gaia_overview, Gaia_EDR3}) for a set of stars identified as probable members of the GD-1 stream (as described in detail in Tavangar et al. in prep). Briefly, this membership model fits the track of the stream --- that is, the mean dependence of stream latitude $\phi_2$, proper motion components in stream-aligned coordinates $\mu_{\phi_1^*}, \mu_{\phi_2}$, and radial velocity $v_{\textrm{rad}}$ (when available) --- using splines with regularly spaced knots in stream longitude $\phi_1$ and variable values in each other coordinate dimension ($\phi_2,  \mu_{\phi_1^*}, \mu_{\phi_2}, v_{\textrm{rad}}$).
Both the mean dependence in each coordinate and the width of the stream as a function of stream longitude are fit using spline tracks, assuming that the intrinsic width of the stream in each component is Gaussian, which allows naturally incorporating measurement uncertainties into this methodology. We use 8 spline knots for each coordinate dimension track, and 5 knots for each width track.
After first filtering \textit{Gaia} sources based on isochrone selection using the best-fitting stellar population parameters from \citet{Price-Whelan_GD1}, the stream track model is fit simultaneously with a model for the stellar background (i.e. field stars) in each coordinate dimension. 
The background stellar distribution is modelled as a mixture of Gaussians for the proper motion components, but is not explicitly modeled in the other phase space coordinates.
From this procedure, each star is assigned a membership probability of belonging to the GD-1 stream; we use stars with a membership probability $>0.3$. We found that using higher probabilities did not greatly change the distributions of the data points.

\subsubsection{Distances}

We use the heliocentric distance measurements reported by \citet{deBoer20} and shown in their Table~1. These were derived using photometry from the Pan-STARRS survey \citep[PS1;][]{Pan-STARRS}, complemented by \textit{Gaia} DR2 astrometry \citep{Gaia_DR2}. The distance measurements are $10^\circ$ apart in $\phi_1$, covering a wide range of $110^\circ$. The values are on the order of $\sim10$~kpc, with uncertainties of $\sim0.08$~kpc. While the above measurements based on the models of the stream's stellar distribution in colour-magnitude space provide reliable relative distances, some uncertainty in the absolute distance scale may remain due to the choice of the isochrones used. In what follows, we compare the absolute distance scale along the streams with stellar standard candles such as Blue Horizontal Branch (BHB) stars and RR Lyrae.

We select BHB stars that are probable members of the GD-1 stream using extinction-corrected Pan-STARRS (PS1) photometry from the \text{Gaia}--PS1 cross-match sample constructed in \citet{Price-Whelan_GD1}. We use an approximate distance track for the stream as a function of stream longitude $\phi_1$ derived from fitting an orbit to the stream to compute PS1 absolute $g$-band magnitudes $M_g$ and de-reddened $g-i$ colors for all stars. We use the stream membership probabilities computed above to select probable stream members based on sky position and astrometry, and then select candidate BHB stars as having $(g-i) < 0$ and $0 < M_g < 1$. Additionally, we select candidate RR Lyrae-type (RRL) stream members using a cross-match of the \textit{Gaia} DR2 RRL sample with a catalog of RRL from the PS1 survey \citep{Sesar:2017}. We again use the astrometric membership probabilities defined above to select probable stream members within this subset. For the BHBs we use the colour-dependant absolute magnitude calibration $M_g(g-r)$ from \citet{LMC_BHB}. For the RR Lyrae, we use the metallicity-dependant absolute magnitude calibration from \citet{RRL_dist} in combination with spectroscopic metallicity [Fe/H]$=-2.3$ reported by \citet{Bonaca_RV}, i.e. for GD-1's RR Lyrae $M_G=0.32$[Fe/H]$+1.11=0.374$. Magnitudes in $G$ band are corrected for the Galactic dust extinction using total reddening values $E(B-V)$ from \citet{SFD} and assuming $A_G=2.27E(B-V)$ \citep[see][]{Iorio_Anatomy}.

We quantify the offset of these new data points by fitting a polynomial to the photometric distance track, and calculating the differences in the distances of the standard candles. The mean offset from the photometric track is less than 0.5~kpc, and only $\sim0.25$~kpc for the RRL. This is less than the scatter of the photometric distances about their fitted polynomial, so we choose to use the unchanged distance values as reported by \citet{deBoer20}.

\subsubsection{Radial velocities}

We use radial velocity measurements from two sources. The first set comes from \citet{Koposov_GD1} and were obtained at the Calar Alto observatory. These consist of radial velocity values of 23 stars which cover a smaller range in $\phi_1$ than the other data, between $-45^\circ$ and $-13^\circ$. Typical uncertainties are $\sim9$~km/s. We also include more precise measurements derived from high-resolution spectroscopy \citep{Bonaca_RV}, which cover a narrower range of the stream between $\phi_1\approx-46^\circ$ and $-29^\circ$. The uncertainties in these values are typically $\sim1$~km/s, giving a tight constraint on the radial velocity track in this range of $\phi_1$.

\subsubsection{Cuts}

While all distance and radial velocity data are included in our likelihood, we choose to use \textit{Gaia} data in the range $-80^\circ<\phi_1<30^\circ$ only for the fitting. This covers a very large extent of sky, and excludes less than $18\%$ of the 4D data points (at $\phi_1<-80^\circ$). This choice was made because the observed $\phi_2$ proper motion tracks significantly change gradient at more negative $\phi_1$, which causes otherwise well-fitting models to have excessively penalised log-likelihoods. For a similar reason, we also exclude \textit{Gaia} stars at $\phi_2>0.9^\circ$. This only removes the stars in the off-track `spur' \citep[see e.g.][]{Price-Whelan_GD1, Bonaca_GD1, deBoer20}. While this spur may have been produced by an interaction with Sgr \citep{deBoer20}, our intention is not to accurately model such an interaction (nor do observational uncertainties allow this). Hence, including the spur would only serve to distort the likelihood of models which are good fits to the main stream track. However, when interpreting the results it must be remembered that GD-1 does show more disruption than we are including in our data. Our data is shown for all 6 dimensions of phase space in Fig.~\ref{fig:data}. The excluded stars in the spur at $\phi_2>0.9^\circ$ are also shown in red.

After these cuts our data consists of 448 stars with positions and proper motions, 12 distance data points and 66 radial velocity measurements.

\subsection{Model setup}

We again use the Lagrange cloud stripping technique to generate models of GD-1. Unlike Pal 5, the identity of the GD-1 progenitor is unknown, and it has likely already dispersed. We therefore choose to linearly decrease the mass of the progenitor from $M_\mathrm{prog}=2\times10^4M_\odot$ at $t=-6$~Gyr to zero at $t=t_\mathrm{disp}$. We consider the scenario where $t_\mathrm{disp}=-3$~Gyr, which roughly correspond to one of the two scenarios for the dispersal of the progenitor suggested by \citet{Webb}. All stars forming such streams are already stripped before the Sgr pericentre at $t\approx-2.7$~Gyr, so will be exposed to any perturbations. We also use their suggestion for the location of the progenitor, placing it at $\phi_1=-40^\circ$. The other five phase space coordinates of the progenitor are allowed to vary, namely $\phi_2$, distance, proper motions in $\phi_1$ and $\phi_2$, and radial velocity.\\

\subsection{Likelihood}\label{section:likelihood}

When comparing our model streams to the data, we wish to consider both the mean stream track and the width. For example, if an interaction with Sgr causes a model stream to significantly widen, its likelihood should be lowered even if its mean track is well-aligned with the data. We therefore use kernel density estimates (KDEs) to approximate the distributions of model stars in bins of $\phi_1$, from which the likelihood of each data point can be calculated \citep[see e.g.][for a similar idea]{Palau_KDE}.

For each model stream, we divide the mock stars into 12 bins of width $\Delta\phi_1=10^\circ$, centred on the distance data points at $\{-85^\circ,-75^\circ, ... , 5^\circ, 15^\circ\}$. To account for the path of the stream through each bin, we estimate its mean track by fitting a quadratic to each of the 5 phase space distributions as functions of $\phi_1$. We use the \texttt{scipy} function \texttt{stats.gaussian\_kde} to fit KDEs to the distributions of mock stars about these quadratics. A popular and rapid algorithm to find the optimal bandwidth of a Gaussian KDE is Silverman's rule of thumb \citep{Silverman}, which works well with unimodal distributions. While most of the mock streams do produce unimodal distributions, we found that perturbations can create low-density secondary streams offset from the main track. These resulted in much larger bandwidths and therefore oversmoothing when using Silverman's rule of thumb. To fix this issue we use only data points within 2 standard deviations of the mean when calculating the bandwidth of each distribution, so far-outlying secondary streams do not contribute. We also do not let the $\phi_2$ and proper motion bandwidths exceed $0.25^\circ$ and 0.05~mas/yr respectively, to prevent oversmoothing. These values are typical widths of the streams in the corresponding dimensions. However, we include all mock stars in each bin when calculating the KDEs themselves. Tests on randomly generated streams showed that this produced KDEs which represented the distributions well. Each of the five dimensions are fitted with their own independent KDE.

We assign a likelihood to a data point by assuming its uncertainty is Gaussian, and multiplying this Gaussian by the KDE. The likelihood of the point is taken as the integral of this product over all space. We ignore the uncertainties in $\phi_2$ (since they are negligible compared to the width of the stream) and do not include the membership probabilities in the likelihood. Finally, we compute the log-likelihood of the model $\mathrm{log}L$ by summing the log-likelihoods of all the data points. The only exceptions to this are when there are too few mock stars in a bin (5 or fewer in our procedure), and when the model stream does not extend over the full range of $\phi_1$ ($-90^\circ<\phi_1<30^\circ$). In both cases we set $\mathrm{log}L=-\infty$ and the model is rejected. We avoid using the normalisation of the density in the likelihood, since this may be sensitive to the time-dependence of the stripping rate from the cluster and would require a more detailed model of the cluster evolution.

\begin{figure*}
    \centering
    \includegraphics[width=\textwidth]{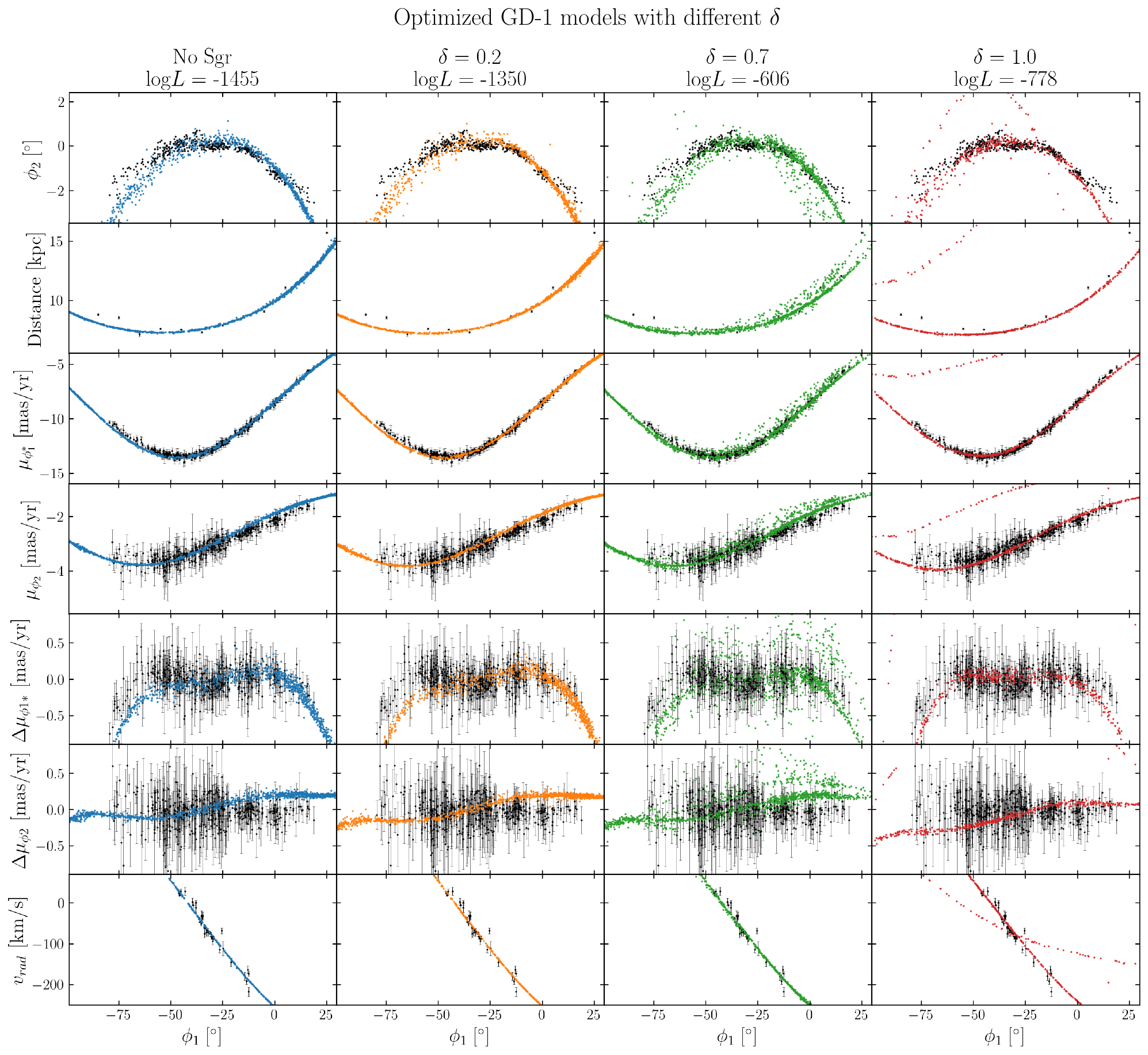}
    \caption{6D tracks of the optimized stream models (coloured) compared to the data (black). From top to bottom, the rows are $\phi_1$, distance, proper motions in $\phi_1$ and $\phi_2$, proper motions relative to a polynomial fitted to the data, and radial velocity. All are plotted against $\phi_1$. Each column corresponds to a different value of $\delta$, with the mass decay rate of Sgr increasing from left to right. The values of log-likelihood for these models are shown above each column.}
    \label{fig:6d_tracks}
\end{figure*}

\subsection{Likelihood optimization}

To find high-likelihood models of the stream for each value of $\delta$, we maximize the log-likelihood function over the 5 phase space coordinates of the progenitor (all except $\phi_1$). We use the \texttt{scipy} function \texttt{optimize.dual\_annealing}, which combines generalized simulated annealing \citep{Xiang} with a local search method. We use the Nelder-Mead simplex algorithm for the local search because the likelihood function contains discontinuities, so we wish to avoid gradient-based methods. \footnote{These discontinuities arise from the discrete nature of our particle-based stream models. For example, a small change in the initial conditions could result in a star moving into a different bin, causing the KDEs of both bins and hence the likelihood to jump discontinuously.} For the optimization with no Sgr, the initial guess was taken from fitting cubic or quadratic polynomials to the data. For the cases with Sgr, we use both this guess and the result of the no Sgr optimization. The optimizer was allowed to explore coordinates in ranges of width $\Delta\phi_2=4^\circ$, $\Delta(\mathrm{distance})=4$~kpc, $\Delta\mu_{\phi_1^*}=\Delta\mu_{\phi_2}=4$~mas/yr, and $\Delta v_\mathrm{rad}=100$~km/s, all much greater than the widths of the observed stream. The optimizations are independently repeated 50 times for no Sgr and 100 times for the models with Sgr (50 for each inital guess), with 10 iterations of the annealing in each run. The local optimization was allowed a maximum of 20 iterations, with a toleration for convergence of 20. A stricter tolerance was not used because the small-scale fluctuations and discontinuities in the log-likelihood would have prevented the optimizer from converging. The solutions were then used as initial guesses for a further 20 repetitions of the optimizer. The optimized model streams from this process for each Sgr model are taken as the results; this produced adequate fits to the data so we do not perform any more iterations. We tested this procedure on mock data generated from models both with and without a massive Sgr. The optimized streams with the correct Sgr models had the highest likelihoods, except that the $\delta=0.2$ model was slightly favoured with data generated with no Sgr. It is therefore difficult to distinguish between different Sgr models of low mass.

\begin{figure*}
    \centering
    \includegraphics[width=\textwidth]{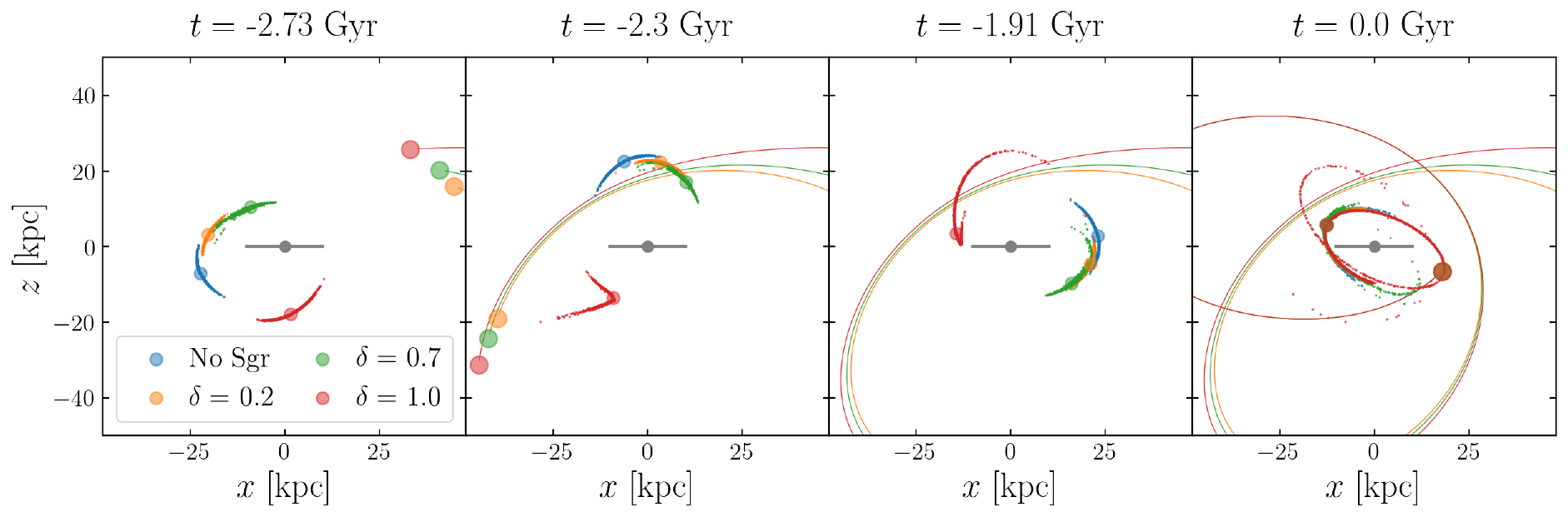}
    \caption{Spatial views of the optimized streams. Each panel shows a different time, while each colour corresponds to a different Sgr model, plotted together for comparison. The projections are edge-on to the MW disc (represented in grey). The coloured lines and large circles show the orbits and positions of Sgr in each model, which are similar over this time period. The $\delta=1.0$ stream is significantly perturbed by Sgr between the first two snapshots. An animated version can be viewed at \url{https://www.youtube.com/playlist?list=PLEleLLhXAwEMx6GcSror-iF-QsskHrXYM}.}
    \label{fig:gal_stills}
\end{figure*}

\begin{figure*}
    \centering
    \includegraphics[width=\textwidth]{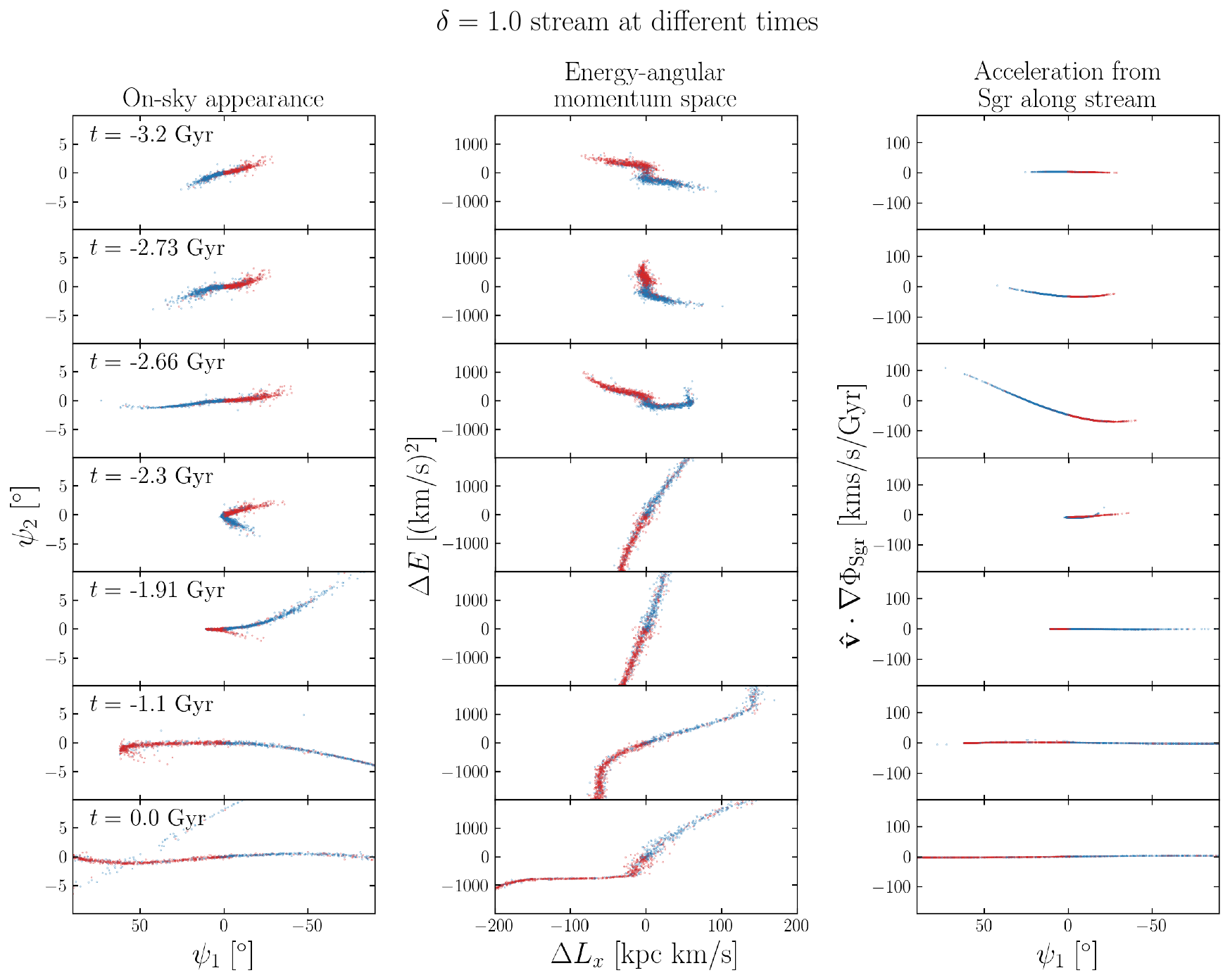}
    \caption{\textbf{Left column}: Spatial appearances of the optimized $\delta=1.0$ stream at a selection of times. ($\psi_1$, $\psi_2$) is the Galactocentric polar coordinate system in which the instantaneous orbital plane of the progenitor is aligned with the $\psi_1$ axis. The progenitor is situated at $\psi_1=\psi_2=0^\circ$ and its motion is in the direction of positive $\psi_1$ (to the left). The stars are colour-coded according to the Lagrange point at which they were released. Blue (red) stars were released from near the inner (outer) Lagrange point, so initially formed the leading (trailing) tail at lower (higher) energy than the progenitor. \textbf{Middle column}: The same stars plotted in energy-angular momentum space, where the origin is shifted to the instantaneous position of the progenitor. \textbf{Right column}: The component of acceleration from Sgr (including MW reflex) along the velocity vector of each particle. An encounter with Sgr at around $t\approx-2.7$~Gyr causes the two tails to be inverted, and the stars which originally formed the trailing tail are now ahead of the leading tail in their orbit. The right-hand column shows that the leading (blue) tail experiences a greater acceleration along the stream than the trailing (red) tail, due to Sgr passing in front of the stream. This causes the leading tail's energy and angular momentum to increase above those of the trailing tail. The spread in energy and angular momentum of both tails also increases significantly as a result of this encounter. An animated version of this figure can be viewed at \url{https://www.youtube.com/playlist?list=PLEleLLhXAwEMx6GcSror-iF-QsskHrXYM}.}
    \label{fig:stills}
\end{figure*}

\subsection{Results}
\label{section:GD1_results}

The optimized model streams are plotted in all 6 dimensions of phase space in Fig.~\ref{fig:6d_tracks}, for several values of $\delta$. The 5th and 6th rows show the proper motions relative to polynomials fitted to the data, to more clearly show details of the proper motion tracks. In Fig.~\ref{fig:gal_stills}, we also show the physical appearance of the streams in galactocentric Cartesian coordinates at four different times, together with the trajectories of Sgr.

Fig.~\ref{fig:6d_tracks} shows reasonably good fits to the data for each of the four models. The two components of proper motion are fitted well in all cases, with the models passing within or close to all the error bars. The distance and radial velocity data are more scattered and scarce, but the model tracks still generally provide good fits. The most obvious deviations from the data are in the on-sky coordinate $\phi_2$. In particular, the mean tracks of the models are offset by $\sim1^\circ$ in $\phi_2$ at $\phi_1<-60^\circ$. This may be because we use a fixed potential; if the potential parameters were allowed to vary it is possible that a better fit may be found. Since the streams are otherwise largely intact, we do not consider that this misalignment rules out a Sgr with these masses.

The $\delta=0.7$ case is one of the most disrupted of all the models, with a secondary stream offset in $\phi_2$, distance and proper motions. This results from a early close encounter with Sgr at $t\approx-4.1$~Gyr, when this Sgr model still has a mass of $5\times10^{10}M_\odot$. However, even this model stream is a reasonable fit to the data for most of $\phi_1$. The GD-1 stream itself exhibits various off-track features \citep{Price-Whelan_GD1}, among them the spur shown in red in Fig.~\ref{fig:data}. There is also the `blob' centred around $\phi_1\approx-15^\circ$. These features were investigated in detail by \citet{deBoer20}, and both have proper motions very similar to the stars in the main stream. Combined with their proximity, this makes them very likely to be related to the main stream. Quantitatively, the proper motion residuals of the two features are $\lesssim2$~mas/yr in magnitude. GD-1 is also accompanied by a secondary stream, known as Kshir \citep{Malhan}, which is offset by $\sim10^\circ$ in $\phi_2$ from GD-1 but shares similar kinematics. The off-track secondary stream in our $\delta=0.7$ model similarly differs by less than 1~mas/yr from the main stream. Hence even though the model stream has had a close encounter with a Sgr of mass $>10^{10}M_\odot$, the level of disruption is not significantly greater than in the observed stream. While the offset in $\phi_2$ is not so great as the observed Kshir stream, there are other examples where Sgr has resulted in differences in $\phi_2$ of $\sim10^\circ$. For example, these are visible in the Pal 1 and BH 176 mock streams in Fig.~\ref{fig:GC_streams}. We have also seen similar cases in other models of GD-1 (not shown here). This suggests that a massive Sgr may be able to produce secondary streams similar to Kshir.

The $\delta=1.0$ stream fits the mean track of the data in all dimensions well, with the exception of a sparse secondary stream. This secondary stream has a large offset in distance and kinematics as well as $\phi_2$, and results from the stream wrapping around the galaxy multiple times (see Fig.~\ref{fig:gal_stills}). It is therefore not comparable to Kshir, unlike the secondary stream in the $\delta=0.7$ model. Despite the good fit of the main track, the stream has not escaped encounters with Sgr. Fig.~\ref{fig:gal_stills} reveals that the stream (in red) is folded at $t\approx-2$~Gyr before extending again in both directions. This stream is also shown in Fig.~\ref{fig:stills} at more snapshots. The left-hand column shows the stream in the galactocentric on-sky coordinates ($\psi_1$, $\psi_2$) as used in Fig.~\ref{fig:GC_streams}. Again the progenitor is always at $\psi_1=\psi_2=0^\circ$ and moving in the direction of positive $\psi_1$. In the middle column energy $E$ is plotted against the x-component of angular momentum $L_x$, where the origin is shifted to the position of the progenitor in $E-L_x$ space. The right-hand column shows the component of acceleration from Sgr along the velocity vector of each particle, against $\psi_1$. The model stars are colour-coded according to the location of their release. Blue stars are released from the inner Lagrange point, so initially form the leading arm, while red stars are from the outer Lagrange point and form the trailing arm.

The left-hand column reveals that the folding and extension of the stream actually results in the arms being inverted; over a period of $\approx1$~Gyr, the leading (blue) arm is overtaken by the progenitor and becomes the trailing arm, and vice versa for the trailing (red) arm. This process is initiated by an encounter with Sgr at $t\approx-2.7$~Gyr; the right-hand column shows an acceleration of the stream particles along their direction of motion. Since Sgr passes in front of the stream, there is a gradient in the magnitude of this acceleration, with the leading tail experiencing the greatest force. The result is that the leading tail gains more energy than the trailing tail, so its new orbit has a longer period and it is overtaken. The middle column also shows a significant increase in the spread of energy and angular momentum.

The changes in energy and angular momentum of stars during an encounter can vary significantly along the stream. For the $\delta=1.0$ stream, some stars (around the centre) experience little change in their orbital parameters. However, stars in the original leading (trailing) tail gained (lost) energy and angular momentum, the latter changing by a factor of up to $\sim1.4$. The resultant gradients in energy and angular momentum along the stream offer a method by which an inverted stream can be detected, if kinematics can be measured precisely enough.

While this reversal happens quite rapidly, this is by no means always the case. Compare the on-sky appearance of the $\delta=1.0$ stream at $t=-1.91$~Gyr to the $\delta=1.0$ Pal 5 stream in Fig.~\ref{fig:GC_streams}. These two streams share the same `folding' of one of the tails, which in the GD-1 model leads to a complete inversion a few hundred Myr later. This folding and consequent bifurcation can therefore be seen as an intermediate stage of inversion, at which one part of the stream is being overtaken by another. This raises the possibility that a stream perturbed by Sgr could be observed undergoing an inversion today, as in our $\delta=1.0$ model of Pal 5. The distance, geometry and relative speed of the encounter are likely the main factors determining the timescale of the inversion. A closer or slower encounter would lead to a larger perturbation, and hence a larger gradient in the orbital frequencies along the stream. However, as the $\delta=1.0$ model suggests, this also results in a more dispersed and hence fainter stream. A more detailed investigation of this process and its results is required to determine the likelihood of observing such a stream, but is beyond the scope of this paper.

The optimized phase space positions themselves show few clear correlations with the Sgr mass. The most obvious trend is in the distance, though even this variation is small. Between the `no Sgr' and $\delta=1.0$ models, the distance decreases monotonically from $7.47$ to $7.27$~kpc. This is accompanied by a change in $\mu_{\phi_1^*}$ from -13.39 to -13.27~mas/yr. These changes likely arise because the optimizer forces the stream to follow an orbit which leads to minimal damage at high Sgr masses. However, this is unlikely to be a general result applicable to the real stream, as such an orbit depends on the details of the Milky Way's potential over time.

To check the sensitivity of our results to the initial mass of the cluster, we generate streams from the optimized phase space coordinates (as in Fig.~\ref{fig:6d_tracks}) using different initial values of $M_\mathrm{prog}$. \citet{Webb} and \citet{Gialluca} have suggested initial masses of roughly $3-6\times10^3M_\odot$, so we test our models with $M_\mathrm{prog}=5\times10^3M_\odot$. The streams generated from the lower mass clusters do not greatly differ from those shown in Fig.~\ref{fig:6d_tracks}. The lower mass streams tend to have their stellar density more concentrated in their centres (near the progenitors). This is expected, since the difference in energy between the Lagrange points is smaller for a lower mass cluster, so the two tails should be less dispersed. Similarly, the sparse secondary stream in the $\delta=1.0$ model has a lower density when generated from the lower mass cluster. However, these differences are generally small and do not affect our conclusions.

%Despite fitting the data reasonably well, this model stream has been completely reversed in its orientation since forming. Such a dramatic encounter may be expected to produce a more obvious signature in the 6D tracks than Fig.~\ref{fig:6d_tracks} would suggest. For example, \citet{deBoer20} showed that GD-1 does not exhibit significant misalignment between its proper motions and on-sky gradient, as seen in the Orphan stream \citep{Erkal_Orphan}. We investigated this by isolating the model stars in the range $-80^\circ<\phi_1<30^\circ$, and found that they only occupy a small fraction of the total length of the stream. During the encounter with Sgr at $t=-2.66$~Gyr, they extend over a range of less than $17^\circ$ in $\phi_1$, or about $20\%$ of the whole stream. The relative compactness of these stars means that the impulse from Sgr does not greatly vary between them, so their subsequent orbits do not substantially diverge from each other.

\subsection{Testing track-proper motion misalignment}
A possible consequence of stream-satellite interactions is a misalignment between the on-sky stream track and proper motions, such that the motion of stars is not along the stream. This has been observed in the Orphan stream, resulting from an interaction with the LMC \citep{Erkal_Orphan}. However, data from \textit{Gaia} DR2 has not shown any inconsistency between the stream track and proper motions of GD-1 \citep{deBoer20}. Here we search for a misalignment in both the EDR3 data and the model streams, to determine whether Sgr is expected to cause this effect. This is particularly relevant for the $\delta=1.0$ model, which appears to fit the data reasonably well despite being inverted. The presence of a clear misalignment in this model could help to rule out the inversion of GD-1.

We estimate the on-sky gradient $\mathrm{d}\phi_2/\mathrm{d}\phi_1$ and its uncertainty by fitting a straight line to stars in overlapping $\phi_1$ bins of width $10^\circ$. The proper motions are first corrected for solar motion, and the ratio $\mu_{\phi_2}/\mu_{\phi_1}$ is calculated without the usual $\mathrm{cos}\phi_2$ factor in the $\phi_1$ proper motion. Correcting for solar motion requires distance estimates of the stars in the data, which we take from a 4th-order polynomial fitted to the distance data. For simplicity we set the fractional uncertainties in distance to 0.1. We then calculate the weighted mean of the proper motion ratio values in each $\phi_1$ bin.

\begin{figure}
    \centering
    \includegraphics[width=\columnwidth]{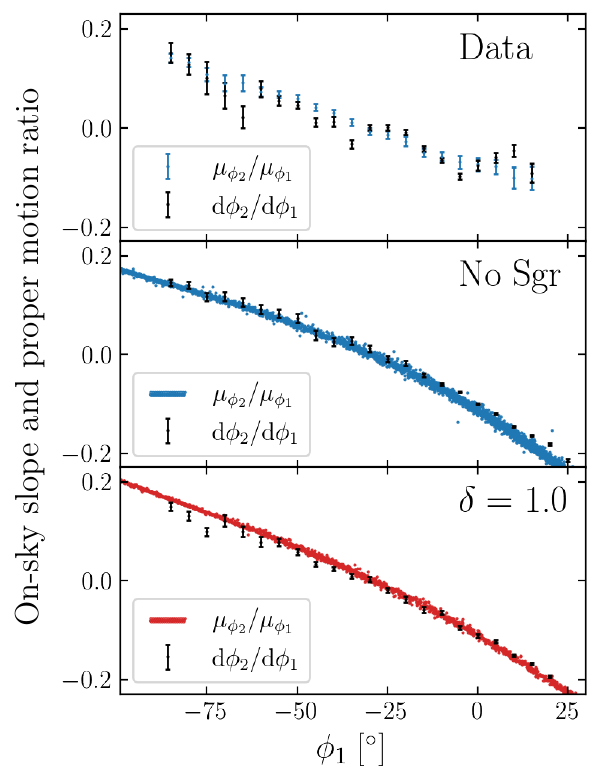}
    \caption{Comparison of on-sky gradient and ratio of proper motions of GD-1, for observations (top panel) and two optimized model streams (lower two panels). With the exception of a few isolated points, the slope and ratio are consistent in both the observations and the models. This is true even for the $\delta=1.0$ stream, which has been perturbed and inverted by Sgr.}
    \label{fig:model_misalignment}
\end{figure}

The on-sky slope and proper motion ratio are plotted in Fig.~\ref{fig:model_misalignment} for the data (top panel), the model stream with no Sgr (middle panel) and the $\delta=1.0$ model (bottom panel). In all cases the slope and proper motion ratio are consistent with each other over the $\phi_1$ range shown, with only a few error bars not overlapping. Hence the more precise EDR3 data support the findings of \citet{deBoer20} that there is no significant misalignment in the observed stream. Neither of the model streams show clear misalignments either. Although the $\delta=1.0$ stream was inverted by a close encounter $\sim2.7$~Gyr ago, in almost all the bins there is very little difference between the slope and proper motion ratio. This may be because of the time elapsed since the encounter; any misalignment is likely to decrease as the stream spreads out along its orbit. With the addition of observational errors in distance and proper motions, any remaining signal could easily be unobservable. Hence the lack of an observed misalignment in GD-1 does not rule out past interactions with Sgr.

\section{Survival of mock GD-1-like streams}\label{section:survival}
\begin{figure*}
    \centering
    \includegraphics[width=\textwidth]{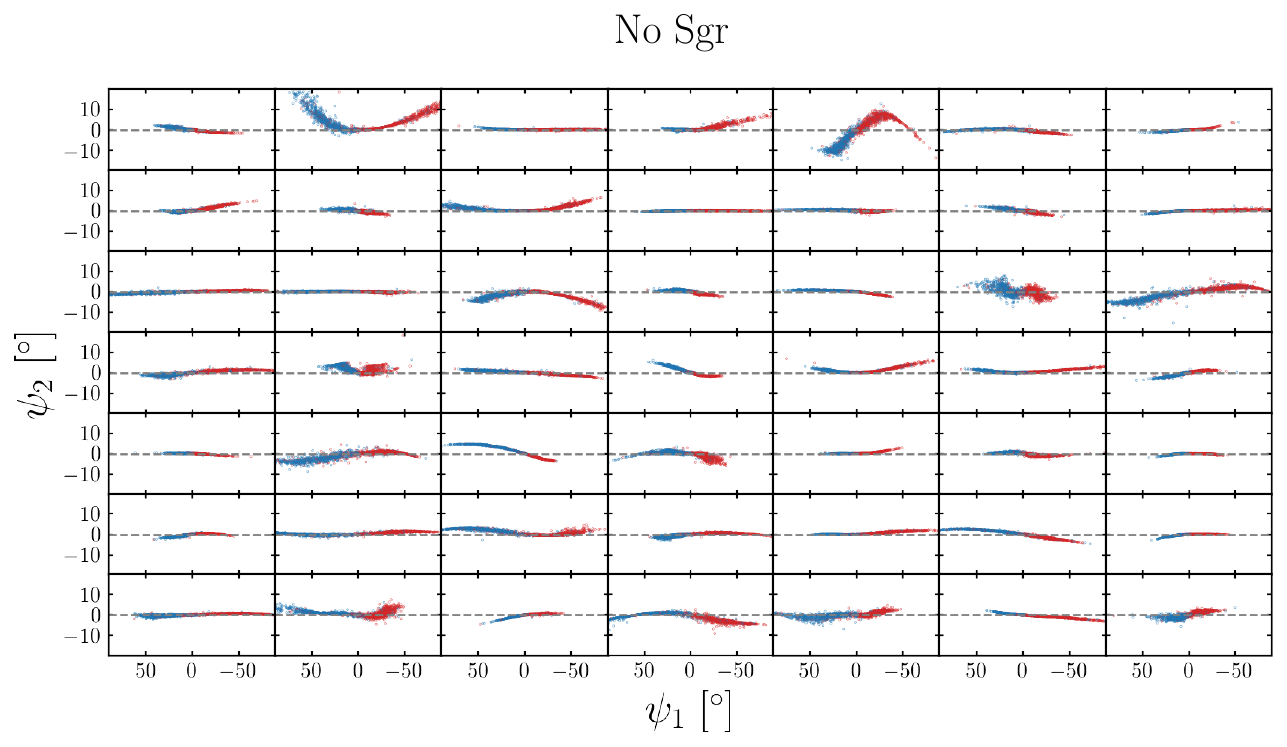}
    \includegraphics[width=\textwidth]{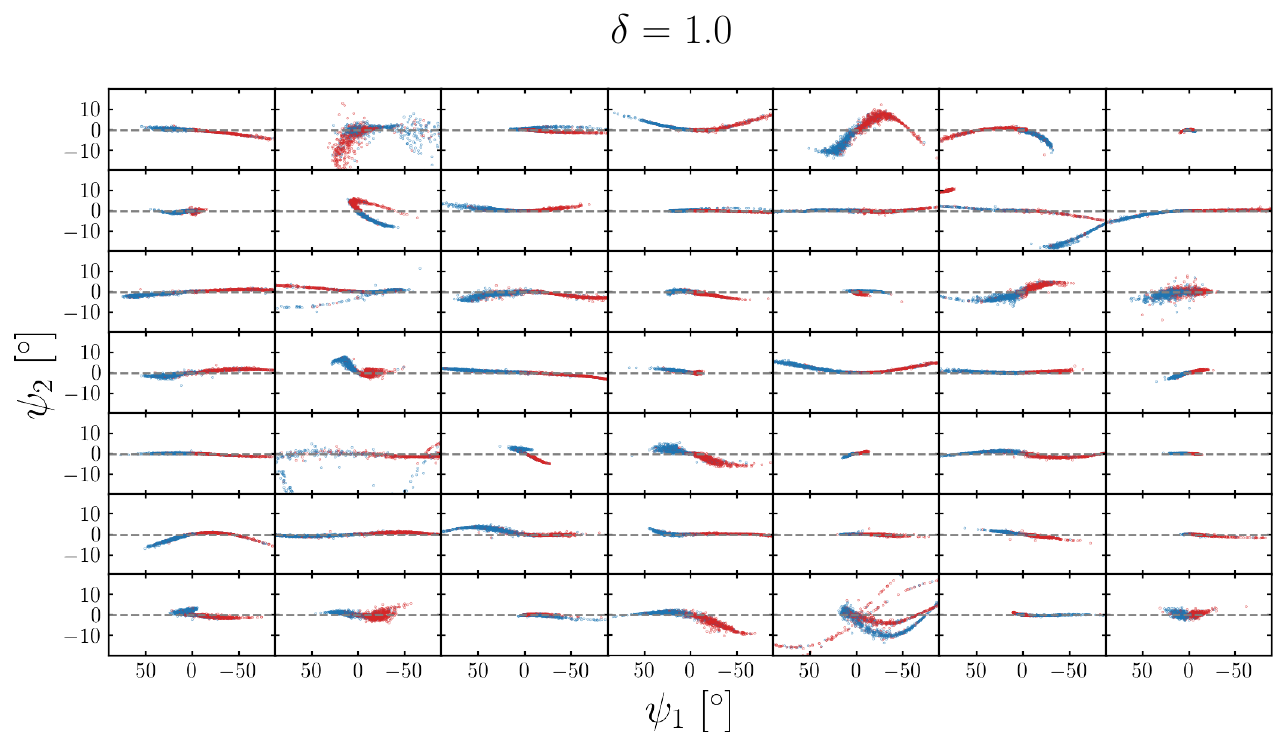}
    \caption{\textbf{Top}: On-sky appearance of 49 mock streams on GD-1-like orbits, with no perturbations from Sgr. The streams are viewed from the galactic centre, and the $\psi_1$ axes (grey dashed lines) are the instantaneous orbital planes of the stream progenitors. As in Fig.~\ref{fig:stills}, the blue (red) stars represent initially leading (trailing) tails. \textbf{Bottom}: the same 49 streams (i.e. generated from progenitors with the same present-day phase space positions), but with perturbations from the $\delta=1.0$ Sgr model. While many streams are largely unchanged, several exhibit disruption from encounters with Sgr. This includes bifurcations resulting from stream `folding' (e.g. top row, third from left), asymmetry in the lengths of the two arms (e.g. second row from bottom, right-most column), and inversion of the arms (e.g. bottom row, second from right).}
    \label{fig:mock_streams}
\end{figure*}

\begin{figure*}
    \centering
    \includegraphics[width=\textwidth]{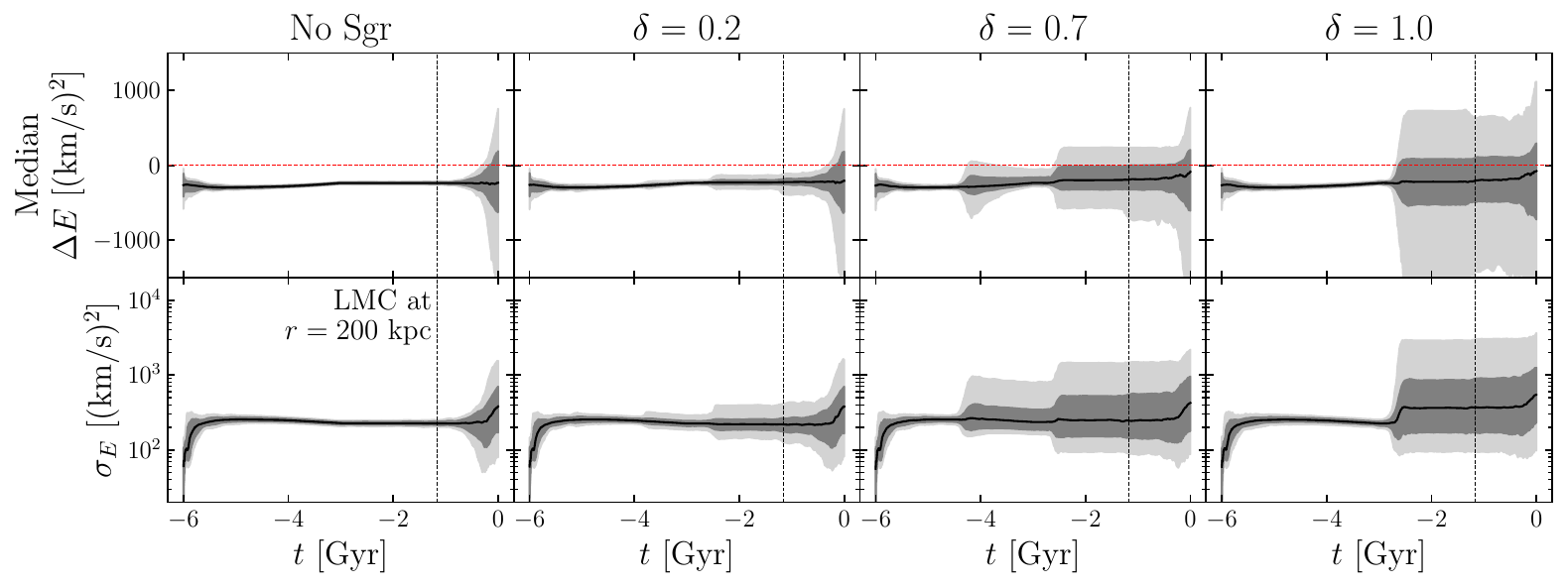}
    \caption{\textbf{Top row}: Median energy of stream stars relative to the progenitor, where each column corresponds to a different model of Sgr (including zero mass on the left). The light and dark grey bands enclose $95\%$ and $68\%$ of streams respectively, while the black line shows the median. The red dashed lines mark the energies of the progenitors. \textbf{Bottom row}: Spread (i.e. standard deviation) of energies of stars in each stream. Both rows show similar behaviour, with the present-day spreads of both values significantly increasing as $\delta$ is increased. A similar effect is visible even without Sgr (left-most column). This is due to the LMC, which crosses $r=200$~kpc at $t\approx-1.1$~Gyr (marked by the vertical black dashed lines).}
    \label{fig:sigma_E}
%\end{figure*}

%\begin{figure*}
    \centering
    \includegraphics[width=0.85\textwidth]{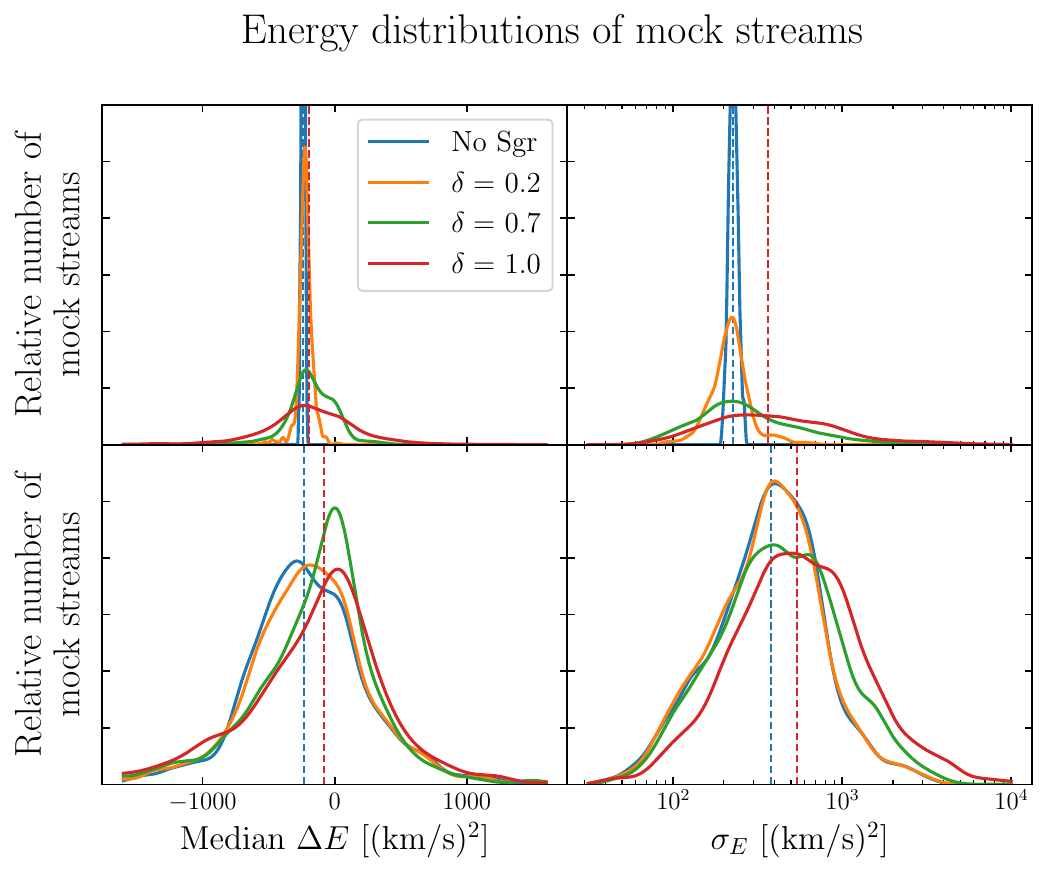}
    \caption{Distributions of $\Delta E$ (left-hand panels)  and $\sigma_E$ (right-hand panel) at $t=-1$~Gyr (top row) and the present-day (bottom row). The colours correspond to different Sgr mass decay profiles, and the vertical dashed lines mark the medians of the distributions for no Sgr and $\delta=1.0$. In the two cases with a massive Sgr ($\delta=0.7$ and $1.0$), the distributions of $\sigma_E$ have tails at higher values, showing that a proportion of the streams become much more spread out in energy. Between $t=-1$~Gyr and the present-day the low-mass distributions spread out due to the effect of the LMC.}
    \label{fig:E_dists}
\end{figure*}

In this section we investigate the effects of a massive Sgr on streams that orbit at similar radii to GD-1, but which are not generally at the present-day location of GD-1. This addresses the probability of such a stream displaying observable signatures of a massive Sgr.

\subsection{Generation of mock GD-1-like streams}

We generate a large number of GD-1-like streams as follows. We integrate the orbit of the maximum likelihood progenitor with no Sgr backwards for 6~Gyr, and select a segment covering exactly 12 radial orbital periods of total length $t_{12}$, from pericentre to pericentre. We find the galactocentric radius $r$, radial velocity $v_r$ and tangential speed $v_t$ as functions of time. For each mock progenitor, we randomly generate a time from a uniform distribution between 0 and $t_{12}$, and assign the present-day radius and radial velocity to be those of GD-1 at the corresponding time. We also set the tangential velocity equal in magnitude to that of GD-1 at that time, but with a direction drawn from a uniform distribution. We also generate the Galactocentric position on the sky from a uniform distribution in solid angle. In this way we create a population of stream progenitors whose orbits have similar radial ranges to GD-1, but which are approximately uniformly distributed in radial phase and orientation. Note however that this cannot be achieved exactly in this non-spherical potential. For a variety of Sgr mass decay parameters $\delta$, we again use the Lagrange Cloud Stripping technique to generate 1000 mock streams from these progenitors. This produces a sample of 1000 different streams for each $\delta$, but each sample shares the same set of present-day progenitor phase space positions.

Since Sgr cannot have passed pericentre with a mass exceeding $\sim10^{10}M_\odot$ in its last two orbits, any stream which formed recently is unlikely to be significantly perturbed by Sgr. Hence we instead study the effects on older streams whose stars were stripped before the Sgr pericentre at $t\approx2.7$~Gyr. Therefore, we set $t_\mathrm{disp}=-3$~Gyr for all the mock streams, so the stars are stripped only in the first 3~Gyr of the simulations.

\subsection{Appearances of mock streams}

The on-sky appearances of 49 mock streams are shown in Fig.~\ref{fig:mock_streams}, with no Sgr (top) and the $\delta=1.0$ model (bottom). The progenitor of each stream has the same present-day phase space position for both Sgr models. These plots reveal a variety of effects that Sgr can have on the appearance of streams. In the most extreme cases one of the tails is folded back on itself (e.g. top row, 3rd column from left), giving an appearance of two streams side by side. Several less perturbed streams instead have a smaller `hook' at the end of one of the tails. There are also examples of inverted streams, as seen in the optimized $\delta=1.0$ GD-1 model (Fig.~\ref{fig:stills}). In these streams, the originally trailing (red) tails now lead the progenitors (i.e. at positive $\psi_1$), and vice versa for the initially leading (blue) tails. Despite the reversals these streams can show little other evidence of perturbations, remaining narrow and lying close to the $\psi_1$ axes (e.g. bottom row, 2nd from right). Observations of such streams may therefore provide no indication that they have been strongly perturbed, depending on the quality of the data. There are also many streams whose appearances are not greatly affected by increasing the mass of Sgr. Some experience nothing more than a small change in the length of one or both tails (e.g. top row, left-most column). It is therefore possible for streams to survive a pericentric passage of a massive Sgr without showing obvious visible signs of an encounter.

\subsection{Energy distributions}

Stellar streams form tightly clustered groups in energy-angular momentum space \citep[e.g.][]{Bonaca_clustering}. One approach to measuring the survival chance of the streams is therefore to measure the spreads of energy and angular momentum. While a stream spread out in energy space does not necessarily mean it will appear disrupted (e.g. see Fig.~\ref{fig:stills}), it remains a strong indication that the stream has been perturbed. Since we are using a triaxial potential, the angular momentum $\mathbf{L}$ is not conserved. However, for orbits in a static MW potential without the LMC or Sgr, the energy $E$ is an integral of motion. We therefore focus on the energy, defined by $E=\frac{1}{2}v^2+\Phi_\mathrm{MW}$, where $\Phi_\mathrm{MW}$ is the static MW potential. We treat the time-dependent parts of the potential induced by the LMC and Sgr as perturbations causing $E$ to vary with time.

In this section, we generate only leading streams; a pair of unperturbed leading and trailing streams has a bimodal energy distribution, which we wish to avoid when studying the spread in energies. Therefore, a large majority of stars in an unperturbed stream are expected to have lower energy than the progenitor from which they were stripped. To see how this is affected by the perturbations, we calculate the difference in energy $\Delta E$ between each star and its progenitor. We also find the standard deviation of the stream star energies $\sigma_E$ for each mock stream, which is a measure of how tightly clustered the stream is in energy space. The median $\Delta E$ and $\sigma_E$ are plotted for each mock stream and for each value of $\delta$ as functions of time in Fig.~\ref{fig:sigma_E}. The present-day distributions are also plotted in Fig.~\ref{fig:E_dists}.

Until about 1~Gyr ago, in the absence of Sgr the median $\Delta E$ and $\sigma_E$ remain almost constant, and there is little variation between the different streams. This is expected, since $E$ is a constant of motion in the static MW potential if the LMC can be ignored. This changes in the last billion years, and the distributions of $\sigma_E$ and median $\Delta E$ become much more spread out. This is due to the LMC perturbing the energy distributions of the streams. The majority become more spread out in energy space (the median $\sigma_E$ increases), but for a significant fraction $\sigma_E$ decreases and the streams become more tightly clustered in energy space.

The same effects are seen when a massive Sgr is introduced. At each pericentric passage of Sgr, streams rapidly become either more or less spread out in enery space over. A minority of these leading streams also have their median energy rise above that of the progenitor; these can be associated with the inverted streams previously discussed, where the progenitor overtakes the leading tail. The top row of Fig.~\ref{fig:E_dists} shows the distributions of $\sigma_E$ and median $\Delta E$ before the LMC begins to affect them. The right-hand panel confirms that the distribution of $\sigma_E$ spreads out when the mass of Sgr is increased. With $\delta=0.7$ or $1.0$, the distribution has a long tail at large $\sigma_E$; these are the streams which become significantly less clustered in energy, and consequently are more likely to disperse spatially. In the last Gyr the LMC erases some of the differences between the distributions, but $\sigma_E$ remains higher on average with the larger values of $\delta$. The median $\sigma_E$ increases by a factor of about 1.4 with $\delta=1.0$ compared to with no Sgr. A massive Sgr therefore only has a mild effect on the average energy spread of streams.

\begin{figure}
    \centering
    \includegraphics[width=\columnwidth]{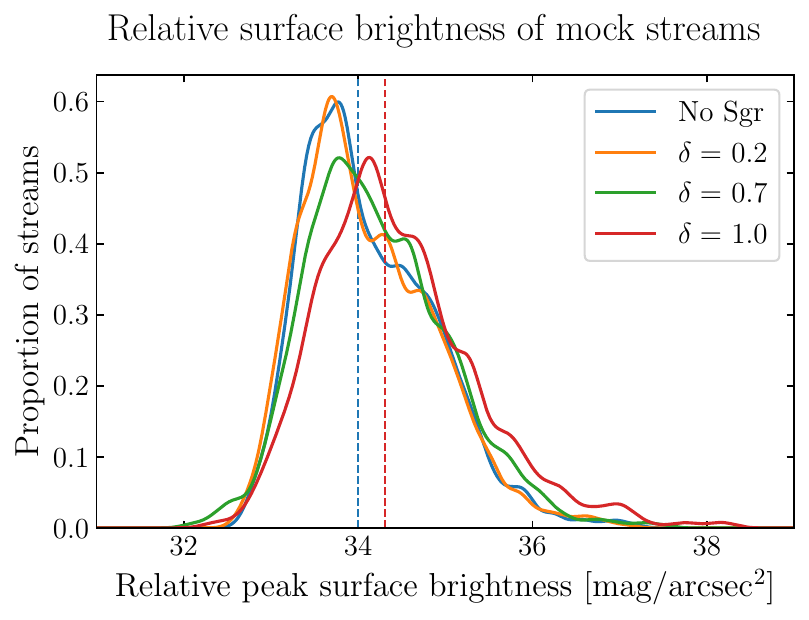}
    \includegraphics[width=\columnwidth]{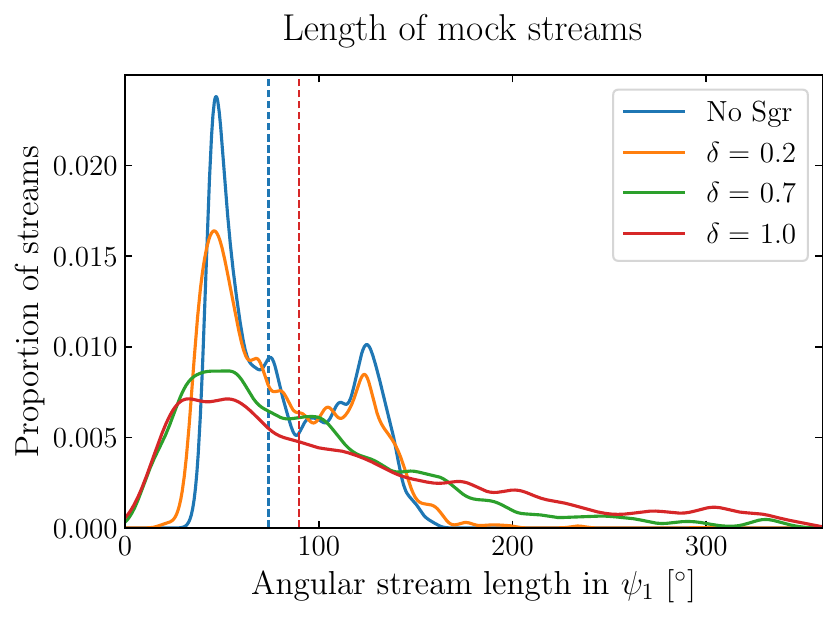}
    \caption{\textbf{Top panel}: Distribution of peak surface brightnesses of the mock streams for several Sgr models. This is only a relative scale, since using different luminosities or numbers of stars in the models would shift all the values. The medians for the two extreme models (no Sgr and $\delta=1.0$) are shown with dashed lines. Higher masses of Sgr result in the peak surface brightness becoming dimmer by about 0.3 mag/arcsec$^2$ on average. \textbf{Bottom panel}: Angular length distributions of mock streams, calculated as the difference between 95th and 5th percentiles of $\psi_1$ distributions. A massive Sgr increases the proportion of streams at both short ($<30^\circ$) and long ($>150^\circ$) lengths. This is likely related to the processes of stream folding and dispersal respectively.}
    \label{fig:peak_dens}
\end{figure}
\subsection{Surface brightness}

The disrupted appearances of many of the streams in Fig.~\ref{fig:mock_streams} due to Sgr suggests that the encounters may affect the surface density of stars in the stream, and hence the surface brightness and probability of detection. While some streams are widened and lengthened with reduced density, others have at least one tail significantly shortened, which may increase the local surface brightness. When studying the dynamics of subhalo encounters with streams, \citet{Erkal_gaps15} similarly found that stream gaps caused by such interactions are associated with caustics of higher density on either side. It is therefore unclear whether the presence of a massive Sgr will increase or decrease the surface brightness and detectability of streams.\\
We consider the relative surface brightness of stars as viewed from the centre of the galaxy. We first divide the mock stars in the range $-60^\circ<\psi_1<60^\circ$ into 12 bins of width $10^\circ$. Reusing the method described in Section~\ref{section:likelihood}, we fit a KDE to the $\psi_2$ distribution in each bin, after subtracting a fitted parabola from the $\psi_2$ values. The contribution from each mock star to the KDE is weighted by its flux as viewed from the galactic centre, using a fiducial luminosity of $1L_\odot$. The peak surface brightness (in units of flux density per solid angle) is taken as the maximum of the KDE, normalised by the total flux from stars in the bin and the bin width. This is then converted to a surface brightness with units of magnitudes per square arcsecond. This is only a relative value, since changing the stripping rate or stellar luminosity would shift each of the values. Distributions of the peak surface brightness are plotted in the top panel of Fig.~\ref{fig:peak_dens} for four Sgr models, with the median values marked with dashed vertical lines for the cases with no Sgr and $\delta=1.0$.

The effect of Sgr on the peak surface brightness is reasonably small. Compared to with no Sgr, the $\delta=1.0$ case only causes the median surface brightness to dim by 0.3 mag/arcsec$^2$, and the distributions have similar widths for all Sgr models. However, the $\delta=1.0$ model does decrease the proportion of streams at high brightness by up to $\sim50\%$. Hence a massive Sgr does disperse the densest parts of some streams, but is unlikely to prevent their discovery entirely.

In the lower panel of Fig.~\ref{fig:peak_dens} we also plot distributions of the streams' angular lengths as viewed from the centre of the galaxy. This is simply defined as the difference between the 95th and 5th percentiles of their $\psi_1$ distributions. In the absence of Sgr virtually all streams have lengths between $30^\circ$ and $160^\circ$ at the present-day. However, introducing a massive Sgr results in more streams with both shorter and longer lengths, anywhere between $0^\circ$ and $360^\circ$. This is unsurprising given the appearances of the streams in Fig.~\ref{fig:mock_streams}; some are significantly shortened while others are more dispersed along the $\psi_1$ axis.
\section{Conclusions}\label{section:conclusions}

We have conducted test particle simulations of stellar streams in the Milky Way under the influence of a Sgr dSph with various initial masses, up to $5\times10^{10}M_\odot$. Each prescribed model of Sgr loses a fixed fraction of its mass at each pericentre, ending with a present-day mass of $4\times10^8M_\odot$. The aim of our analysis is to address whether such high masses are compatible with the survival of streams formed over 3~Gyr ago, with a focus on the well-studied GD-1 stream. Our principal conclusions are summarised below.

\begin{enumerate}[label=\textbf{(\roman*)}]
    \item \textbf{A Sgr with an initial mass of $\gtrsim10^{10}M_\odot$ is able to cause significant disruption to old stellar streams in the inner halo of the Milky Way.} As examples we have generated mock stellar streams from three known globular clusters, and shown that in each case the stream becomes significantly more perturbed when the initial mass of Sgr is raised to  $4\times10^{10}M_\odot$ or more. However, this only applies to the stars stripped more than $\sim2.7$~Gyr before the present-day. This is because realistic models of Sgr require at least two orbital periods to lose mass from $\sim10^{10}M_\odot$ to the present-day value of $4\times10^8M_\odot$. Encounters with a massive Sgr can therefore occur no more recently than around the pericentre two orbital periods ago. In our choice of potential this pericentre occurs about $2.7$~Gyr before the present, though this time will be shorter for some other realistic and commonly used potentials \citep[e.g.][]{McMillan17}. Any stream formed since that time is able to grow unperturbed, and the effects of Sgr can be safely ignored.

    \item \textbf{Pal 5 tidal tails.} We have generated models of the Pal 5 tidal tails that qualitatively reproduce features in the observed stream, in particular the asymmetry between the leading and trailing tails. This is produced by an interaction with Sgr of mass $>10^9M_\odot$, which causes one of the tails to be folded back on itself and shortened. While the simulated shortened tails tend to be longer than the observed one, we find that this is sensitive to the model setup. An interaction with Sgr therefore remains a possible candidate for the origin of the Pal 5 asymmetry.
    
    \item \textbf{Fitting the GD-1 stream with different models of Sgr.} For several different mass decay profiles of Sgr, we fitted models to 6D data of the GD-1 stream using a kernel density estimate to define the likelihood. For each mass of Sgr, the optimized model fitted the data reasonably well, with the most obvious deviations being in the on-sky coordinate $\phi_2$. For the highest initial mass ($5\times10^{10}M_\odot$, the $\delta=0.7$ model), an encounter with Sgr creates off-track features parallel to the main track. The offsets of these features from the main stream are comparable to those of the `spur' and `blob' in the observed GD-1 stream, so the degree of disruption induced by this mass of Sgr is consistent with observations. Hence, we conclude that a Sgr of mass $\sim5\times10^{10}M_\odot$ as predicted by \citet{Read} and \citet{Laporte18} is compatible with the current state of the GD-1 stream, even if the stream was formed more than 3~Gyr ago.
    
    \item \textbf{Stream reversals.} One of the highest-likelihood GD-1 models corresponds to the fastest-decaying Sgr model ($\delta=1.0$, initial mass $4\times10^{10}M_\odot$). The mean track of the mock stars fits the data reasonably well in all dimensions of phase space. Interestingly however, the stream is inverted with respect to its original orientation. The stars released from the inner Lagrange point which originally formed the leading tail now form the trailing tail, behind the stars released from the outer Lagrange point. This reversal is initiated by an encounter with Sgr and happens over a period of $\sim1$~Gyr. It results from Sgr passing in front of the stream, causing the leading stars to move onto orbits with slightly lower frequencies. They are subsequently overtaken by the trailing stars over the next orbital periods, accompanied by a contraction of the stream. The stream then extends again along its orbit, with the originally trailing stars now leading. We find that neither the data nor this stream shows any significant misalignment between the on-sky slope and proper motion ratio, so such a scenario is difficult to rule out without more precise data. It is notable that such a dramatic change could take place and leave such a well-fitting stream.
    
    \item \textbf{Mock streams.} We generated 1000 mock streams with apocentric and pericentric distances similar to those of GD-1, but with randomized orbital planes and phases. We set the dispersal time of each cluster to $t_\mathrm{disp}=-3$~Gyr so that all stars are susceptible to interactions with Sgr around the pericentre at $t\approx-2.7$~Gyr. The massive Sgr models create a wide range of visible features in the streams, including asymmetry between the lengths of the two tails, bifurcations induced by `folding' of a tail, and reversals of the leading and trailing tails.
    
    \item \textbf{Energy distributions of mock streams.} We calculated the energy of stars in each mock streams, using only the leading tails (i.e. stars stripped from the inner Lagrange point). We found their median offset in energy $\Delta E$ from that of the progenitor, and the standard deviations of their energies $\sigma_E$. At $t=-1$~Gyr, a massive Sgr has caused the spread in energy $\sigma_E$ to increase for most streams, by up to an order of magnitude. However, the distributions of $\sigma_E$ for each Sgr model are more similar at the present-day, due to the influence of the LMC in the last Gyr. With a Sgr of initial mass $4\times10^{10}M_\odot$ (the $\delta=1.0$ model), the present-day median value of $\sigma_E$ is also increased by only a factor of 1.4 compared to the case with no Sgr. The present-day clustering of stream stars in energy space is therefore a poor indicator of the initial mass of Sgr.
    
    \item \textbf{Surface brightness and length of mock streams.} We estimated the peak surface brightness of stars in each of the same 1000 mock streams as viewed from the centre of the galaxy. The $\delta=1.0$ Sgr model (with initial mass $4\times10^{10}M_\odot$) reduces (dims) the median peak surface brightness by about 0.3 mag/arcsec$^2$ compared to the model with no Sgr. We conclude that a massive Sgr only slightly reduces the chance of detecting a stream formed before $t=-3$~Gyr, so mass estimates of $\sim5\times10^{10}M_\odot$ are compatible with observed streams being more than 3~Gyr old. The massive Sgr models also result in a much wider range of stream lengths than with no Sgr.
\end{enumerate}

We have demonstrated that a Sgr of mass $\sim10^{10}M_\odot$ is capable of causing significant disruption to stellar streams formed more than 3~Gyr ago, including folding and reversal of the arms. However, these large masses are compatible with the survival of the streams, and it is possible to recreate GD-1 with no more damage than is observed. Our results suggest that the influence of Sgr should be considered when studying the perturbations of streams in the Milky Way, as encounters over $2.5$~Gyr could result in present-day disruption. Being one of the largest satellites of the Milky Way, knowing the former mass of Sgr is key to reconstructing its history. As new data from \textit{Gaia} and other surveys becomes available, stellar streams will play a crucial role in understanding our galaxy's past.

\section*{Acknowledgements}

We thank the anonymous referee whose comments have helped to improve this article. AMD thanks the Science and Technology Facilities Council (STFC) for a PhD studentship. We are grateful to Eugene Vasiliev and the Cambridge Streams and CCA Dynamics groups, for helpful discussions and suggestions.

This work has made use of data from the European Space Agency (ESA) mission
{\it Gaia} (\url{https://www.cosmos.esa.int/gaia}), processed by the {\it Gaia}
Data Processing and Analysis Consortium (DPAC,
\url{https://www.cosmos.esa.int/web/gaia/dpac/consortium}). Funding for the DPAC
has been provided by national institutions, in particular the institutions
participating in the {\it Gaia} Multilateral Agreement.

This research made use of Astropy,\footnote{http://www.astropy.org} a community-developed core Python package for Astronomy \citep{astropy:2013, astropy:2018}.

This work was funded by UKRI grant 2604986. For the purpose of open access, the author has applied a Creative Commons Attribution (CC BY) licence to any Author Accepted Manuscript version arising.

%%%%%%%%%%%%%%%%%%%%%%%%%%%%%%%%%%%%%%%%%%%%%%%%%%
\section*{Data Availability}

The data containing probable GD-1 members from \textit{Gaia} EDR3 will be published with a future paper (Tavangar et al. in prep). The RRL and BHB data are provided alongside sample scripts at \url{https://doi.org/10.5281/zenodo.7037883}.

%%%%%%%%%%%%%%%%%%%% REFERENCES %%%%%%%%%%%%%%%%%%

% The best way to enter references is to use BibTeX:

\bibliographystyle{mnras}
\bibliography{bibliography}

%%%%%%%%%%%%%%%%%%%%%%%%%%%%%%%%%%%%%%%%%%%%%%%%%%

%%%%%%%%%%%%%%%%% APPENDICES %%%%%%%%%%%%%%%%%%%%%

\appendix
\section{Milky Way Potential}\label{section:appendix_MW}
Here we describe the triaxial MW potential used throughout this paper. This is identical to the triaxial fiducial model used by \citet{tango} to study the Sgr stream.

The bulge is spherical, with total mass $1.2\times10^{10}M_\odot$ and a density profile
\begin{equation}
    \rho_\mathrm{b}\propto(1+r/r_\mathrm{b})^{-1.8}\,\mathrm{exp}[-(r/u_\mathrm{b})^2],
\end{equation}
where the scale radius is $r_\mathrm{b}=0.2$~kpc, and the cutoff radius is $u_\mathrm{b}=1.8$~kpc.

The disc has total mass $5\times10^{10}M_\odot$, and has density
\begin{equation}
    \rho_\mathrm{d}\propto\mathrm{exp}[-R/R_\mathrm{d}]\,\mathrm{sech}^2(z/2h_\mathrm{d}),
\end{equation}
where the scale radius is $R_\mathrm{d}=3$~kpc and the scale height is $h_\mathrm{d}=0.4$~kpc.

The halo is triaxial, with a radius-dependent shape and orientation. The inner halo's density distribution is oblate with an axis ratio of $q_{in}=0.64$, and its minor axis is aligned with the $z$-axis. Beyond a shape transition radius of $r_q=54$~kpc, the halo is prolate with axis ratios $1.45:1.37:1$. The major $Z$-axis is aligned with the $z$-axis, while the minor and intermediate axes $X$ and $Y$ lie in the galactic plane. However, the $X-Y$ axes are rotated by an angle $\alpha_q=-25^\circ$ from the $x-y$ axes.

The halo density profile is given by \citep{tango}:
\begin{equation}
    \rho_\mathrm{h}(s)\propto(s/r_\mathrm{h})^{-\gamma}[1+(s/r_\mathrm{h})^\alpha]^{(\gamma-\beta)/\alpha}\mathrm{exp}[-(s/u_\mathrm{h})^\eta],\\
\end{equation}
where $s$ is the `sphericalized radius'. This can be considered as the magnitude of a position vector appropriately scaled by the axis ratios. The density law itself is a \citet{Zhou} profile, modulated by an exponential cut-off which ensures a finite total mass. We use a scale radius $r_\mathrm{h}=7.36$~kpc, inner slope $\gamma=1.2$, outer slope $\beta=2.4$, and transition steepness $\alpha=2.4$. The outer cutoff radius is $u_\mathrm{h}=200$~kpc and the exponential steepness is $\eta=2$. The halo density is normalised to give a circular velocity of approximately 235~km/s at the Solar radius.

\section{Calculating the orbit of a massive Sgr}\label{section:Sgr_orbit}

Below we briefly explain the procedure for obtaining our prescribed mass decay profile for Sgr (described in Section~\ref{section:setup_Sgr}), while maintaining consistency with Chandrasekhar dynamical friction (equation~\ref{eq:DF}). We first integrate the orbit of Sgr from its present-day position back to the penultimate pericentre. We include the influence of dynamical friction using the present-day mass $M_\mathrm{Sgr}(0)$. We record the times of this pericentre and subsequent apocentre, and calculate the necessary exponential mass decay rate such that $M_\mathrm{Sgr}$ decreases by a factor of $10^{-\delta}$ between the pericentre and apocentre. We next re-integrate the orbit using the newly computed mass decay profile prior to the apocentre. This process is repeated with the previous orbits until we have accounted for all pericentre passages occurring after $t_\mathrm{start}$. We note that this does not provide exact consistency with equation~\ref{eq:DF}, since the pericentre times change slightly when the orbit is re-integrated with a different mass. However, we found that in practice these changes were extremely small and have a negligible effect on the orbit and $M_\mathrm{Sgr}(t)$.

We treat Sgr as a test particle (with dynamical friction) when integrating its orbit, even when it has a non-negligible mass. Below is a brief justification for this approach. If both the MW and Sgr are treated as rigid but with time-varying masses, their equations of motion in an inertial frame can be approximated as
\begin{equation}\label{eq:Sgr_2body}
\begin{split}
\ddot{\mathbf{x}}_\mathrm{MW}&=-\nabla\Phi_\mathrm{Sgr}(-\mathbf{r}_\mathrm{Sgr}, t)\\
\ddot{\mathbf{x}}_\mathrm{Sgr}&=-\nabla\Phi_\mathrm{MW}(\mathbf{r}_\mathrm{Sgr}, t)+\mathbf{a}_\mathrm{DF}\\
\mathbf{r}_\mathrm{Sgr}&\equiv\mathbf{x}_\mathrm{Sgr}-\mathbf{x}_\mathrm{MW},
\end{split}
\end{equation}
where the LMC is ignored since we are considering early times, when it is at large radii. The position vector of Sgr in the non-inertial galactocentric frame $\mathbf{r}_\mathrm{Sgr}$ therefore evolves according to
\begin{equation}
\begin{split}
    \ddot{\mathbf{r}}_\mathrm{Sgr}&=-\nabla\Phi_\mathrm{MW}(\mathbf{r}_\mathrm{Sgr},t)+\nabla\Phi_\mathrm{Sgr}(-\mathbf{r}_\mathrm{Sgr},t)+\mathbf{a}_\mathrm{DF}\\
    &\approx-\nabla\Phi_\mathrm{MW}(\mathbf{r}_\mathrm{Sgr},0)+\mathbf{a}_\mathrm{DF}\\&\quad-\nabla[\Phi_\mathrm{MW}(\mathbf{r}_\mathrm{Sgr},t)-\Phi_\mathrm{MW}(\mathbf{r}_\mathrm{Sgr},0)]\\&\quad+\nabla[\Phi_\mathrm{Sgr}(-\mathbf{r}_\mathrm{Sgr},t)-\Phi_\mathrm{Sgr}(-\mathbf{r}_\mathrm{Sgr},0)]
    %&\sim-\frac{G[M_\mathrm{MW}(r_\mathrm{Sgr}, t)+M_\mathrm{Sgr}(t)]}{r_\mathrm{Sgr}^3}\,\mathbf{r}_\mathrm{Sgr},
\end{split}
\end{equation}
where we have divided the MW potential into constant and time-varying parts, and the present-day potential of Sgr has been partially discarded since it is negligible compared to the other terms.

Meanwhile, in our models we treat Sgr as a test particle orbiting a fixed, time-independent MW. In this case the equation of motion is
\begin{equation}
    \ddot{\mathbf{r}}_\mathrm{Sgr}=-\nabla\Phi_\mathrm{MW}(\mathbf{r}_\mathrm{Sgr},0)+\mathbf{a}_\mathrm{DF}
\end{equation}
We see that the two situations are equivalent if $\nabla[\Phi_\mathrm{MW}(\mathbf{r}_\mathrm{Sgr},t)-\Phi_\mathrm{MW}(\mathbf{r}_\mathrm{Sgr},0)]=\nabla[\Phi_\mathrm{Sgr}(-\mathbf{r}_\mathrm{Sgr},t)-\Phi_\mathrm{Sgr}(-\mathbf{r}_\mathrm{Sgr},0)]$. Since the potential of Sgr is spherical, this requires that
\begin{equation}
    \Phi_\mathrm{MW}(\mathbf{r}_\mathrm{Sgr},0)-\Phi_\mathrm{MW}(\mathbf{r}_\mathrm{Sgr},t)=-\frac{G[M_\mathrm{Sgr}(r_\mathrm{Sgr},t)-M_\mathrm{Sgr}(r_\mathrm{Sgr},0)]}{r_\mathrm{Sgr}},
\end{equation}
where $M_\mathrm{Sgr}(r_\mathrm{Sgr},t)$ is the mass of Sgr enclosed within a radius of $r_\mathrm{Sgr}\equiv|\mathbf{r}_\mathrm{Sgr}|$ at time $t$.

Treating Sgr as a test particle orbiting a static MW is therefore equivalent to assuming that the mass lost by Sgr between time $t$ and $t=0$ is gained by the MW, and is distributed in a sphere of radius $<r_\mathrm{Sgr}$. This is not an unreasonable assumption, since much of the mass will be stripped close to the pericentres. Note that using equations~\ref{eq:Sgr_2body} would necessarily involve a time-varying MW mass, which would also require an assumption about the distribution of mass stripped from Sgr. Since all such models inevitably contain many assumptions about the potentials and the effect of dynamical friction, we consider our simplified models sufficient for our purposes.

The orbits and mass loss profiles of our Sgr models over the last 2 Gyr can be compared to the simulations of \citet{last_breath} and \citet{tango}. An advantage of our prescription is that dynamical friction does not have a significant influence on the orbit during this period, because of the low mass. Hence, each of our Sgr models follows a similar orbit over the last two periods (see Figs.~\ref{fig:Sgr_mass_orbits} and \ref{fig:gal_stills}). These orbits are also good matches to those of \citet{tango} (see their Figs.~6 and 9), which were shown to reproduce the Sgr stream well.

Our mass loss profiles are also reasonable matches to the best-fitting simulations of the Sgr remnant by \citet{last_breath} (see their Fig.~9). At the present-day their total masses are each around $4\times10^8M_\odot$, having decreased from $1-4\times10^9M_\odot$ in the last 2~Gyr. During the same period our models decay from roughly $0.6-4\times10^9M_\odot$ to $4\times10^8M_\odot$ at the present-day. In summary, our prescribed models of Sgr compare well with some of the most recent and successful simulations of the Sgr remnant and stream.

%%%%%%%%%%%%%%%%%%%%%%%%%%%%%%%%%%%%%%%%%%%%%%%%%%

% Don't change these lines
\bsp	% typesetting comment
\label{lastpage}
\end{document}